\documentclass{aa} 
\usepackage{amsmath,amstext}
\usepackage{balance}
\usepackage{color}
\usepackage{comment}
\usepackage{multirow}
\usepackage{hyperref}

\newcommand{\src}{GRO~J1750-27 }
\graphicspath{{./}{figures/}}

\begin{document}

   \title{The unaltered pulsar: GRO J1750-27, a supercritical X-ray neutron star that does not blink an eye}



   \author{C.~Malacaria
   \inst{1} 
   \and L.~Ducci\inst{2} 
   \and M.~Falanga\inst{1} 
   \and D.~Altamirano\inst{3} 
   \and E.~Bozzo\inst{4}
   \and S.~Guillot\inst{5}
   \and G.~K.~Jaisawal\inst{6} 
   \and P.~Kretschmar\inst{7} 
   \and M.~Ng\inst{8} 
   \and P.~Pradhan\inst{9} 
   \and R.~Rothschild\inst{10} 
   \and A.~Sanna\inst{11} 
   \and P.~Thalhammer\inst{12} 
   \and J.~Wilms\inst{12}
   } 
   
   \institute{International Space Science Institute (ISSI), Hallerstrasse 6, 3012 Bern, Switzerland\\
   \email{cmalacaria.astro@gmail.com}
   \and
   Institut f\"ur Astronomie und Astrophysik, Kepler Center for Astro and Particle Physics, Universit\"at T\"ubingen, Sand 1, 72076 T\"ubingen, Germany
   \and
   School of Physics and Astronomy, University of Southampton, Southampton, Hampshire SO17 1BJ, UK
   \and
   Department of Astronomy, University of Geneva, Chemin d'Ecogia 16, 1290, Versoix, Switzerland
   \and Institut de Recherche en Astrophysique et Plan\'{e}tologie, UPS-OMP, CNRS, CNES, 9 avenue du Colonel Roche, BP 44346, F-31028 Toulouse Cedex 4, France
   \and DTU Space, Technical University of Denmark, Elektrovej 327-328, DK-2800 Lyngby, Denmark
   \and
   European Space Agency (ESA), European Space Astronomy Centre (ESAC), Camino Bajo del Castillo s/n, 28692 Villanueva de la Cañada, Madrid, Spain
   \and MIT Kavli Institute for Astrophysics and Space Research, Massachusetts Institute of Technology, Cambridge, MA 02139, USA
   \and Embry Riddle Aeronautical University, Department of Physics Prescott Campus, 3700 Willow Creek Road, Prescott, AZ 86301
   \and Center for Astrophysics and Space Sciences, University of California San Diego, La Jolla, California 92093
   \and Dipartimento di Fisica, Universit\`a degli Studi di Cagliari, SP Monserrato-Sestu km 0.7, 09042 Monserrato, Italy
   \and Remeis-Observatory and Erlangen Centre for Astroparticle Physics, Friedrich-Alexander-Universit\"at Erlangen-N\"urnberg, Sternwartstr.~7, 96049 Bamberg, Germany
   }

\abstract
{
When accreting X-ray pulsars (XRPs) undergo bright X-ray outbursts, their luminosity-dependent spectral and timing features can be analyzed in detail.
The XRP GRO J1750-27 recently underwent one such episode, during which it was observed with \textit{NuSTAR} and monitored with \textit{NICER}.
Such a data set is rarely available, as it samples the outburst over more than 1 month at a luminosity that is always exceeding ${\sim}5\times10^{37}\,$erg/s. This value is larger than the typical critical luminosity value, where a radiative shock is formed above the surface of the neutron star.
Our data analysis of the joint spectra 
returns a highly ($N_H\sim(5-8)\times10^{22}\,$cm$^{-2}$) absorbed spectrum showing a K$\alpha$ iron line, a soft blackbody component likely originating from the inner edge of the accretion disk, and
confirms the discovery of one of the deepest cyclotron lines ever observed, at a centroid energy of ${\sim}44\,$keV corresponding to a magnetic field strength of $4.7\times10^{12}\,$G.
This value is independently supported by the best-fit physical model for spectral formation in accreting XRPs which, in agreement with recent findings, favors a distance of 14 kpc and also reflects a bulk-Comptonization-dominated accretion flow.
Contrary to theoretical expectations and observational evidence from other similar sources, the pulse profiles as observed by \textit{NICER} remain remarkably steady through the outburst rise, peak and decay.
The \textit{NICER} spectrum, including the iron K$\alpha$ line best-fit parameters, also remain almost unchanged at all probed outburst stages, similar to the pulsed fraction behavior.
We argue that all these phenomena are linked and interpret them as resulting from a saturation effect of the emission from the accretion column, which occurs in the high-luminosity regime.}


 \titlerunning{NICER and NuSTAR observations of \src}
 \authorrunning{Malacaria et al.}
\maketitle

\section{Introduction}\label{sec:introduction}

Accreting X-ray pulsars (XRPs) are binary systems consisting of a neutron star (NS) and a donor companion star.
In these systems, the NS can accrete matter supplied by the companion either via stellar wind or Roche-lobe overflow, and therefore emit in the X-ray domain.
Most of these systems pertain to the subclass of Be/X-ray binaries (BeXRBs), in which the donor star is of B spectral type, expelling its wind under the form of a circumstellar decretion disk, characterized by H$\alpha$ Balmer emission lines.
For recent reviews of such systems see, e.g., \citet[]{Malacaria20} and \citealt{Mushtukov22}.

\src is an XRP discovered with the BATSE observatory \citep[]{Wilson+95}, which also detected pulsations at about 4.4 s and an orbital period of about 30 days \citep{Scott97}.
More recently, the spin period was measured at ${\sim}4.45\,$s \citep[]{Shaw09, Devaraj22}. 
Although no optical counterpart has yet been unambiguously identified, the X-ray timing behavior \citep{Scott97} and the infrared spectral properties \citep[]{Lutovinov19} hint at a BeXRB system at a distance of between 14 and 22\,kpc.
The closest Gaia counterpart in the Early Data Release 3 \citep[EDR3]{Fabricius21} is found at $2.8\,$arcsec from the \textit{Chandra} source position determined by \citet[]{Lutovinov19} and outside of its 
90\% confidence level (c.l.) ellipse region, thus the two sources are likely not associated.
In addition, parallax nor photogeometric distance has been measured for the Gaia counterpart \citep[]{Bailer-Jones21}. 
More recently, \citet[]{Sharma22} used accretion torque modeling to further constrain the distance to 13.6-16.4 kpc.

During the BATSE era, the only outburst observed from \src was the same that led to its discovery, and only a few additional outbursts have been observed by the currently operating all-sky monitors \citep[]{Krimm08,Finger+Wilson14, Boissay15}. The most recent outburst was observed in 2021 \citep[]{Malacaria21_AtelGRO}.
In 2008, the source exhibited an outburst that was monitored with \textit{INTEGRAL} and \textit{Swift}/XRT, and peaked at about 250 mCrab \citep[]{Shaw09}. 
The \textit{INTEGRAL} ISGRI and JEM-X1 joint spectra ($3-50\,$keV) at the peak of the outburst were fit with a cutoff power-law ($\Gamma=-0.15, E_{cut}=6.0\,$keV) or a Comptonization model (\texttt{CompTT}, \citealt[]{Titarchuk94}, with plasma temperature kT$_{e}=4.6\,$keV, and plasma optical depth $\tau=6.4$).

During the 2021 outburst, we initiated a monitoring campaign with \textit{NICER} and \textit{NuSTAR}. Preliminary results from those observations allowed to identify an electron cyclotron resonant scattering feature (CRSF) for the first time in this source at ${\sim43}\,$keV \citep[]{MalacariaAtel2022, Devaraj22, Sharma22}.
Here we further investigate the implications of the spectral analysis carried out with phenomenological and physical models, and take advantage of NICER data to explore the timing behavior and pulse-profile evolution of the source throughout the outburst.

\begin{figure}[!t]
\includegraphics[width=.47\textwidth]{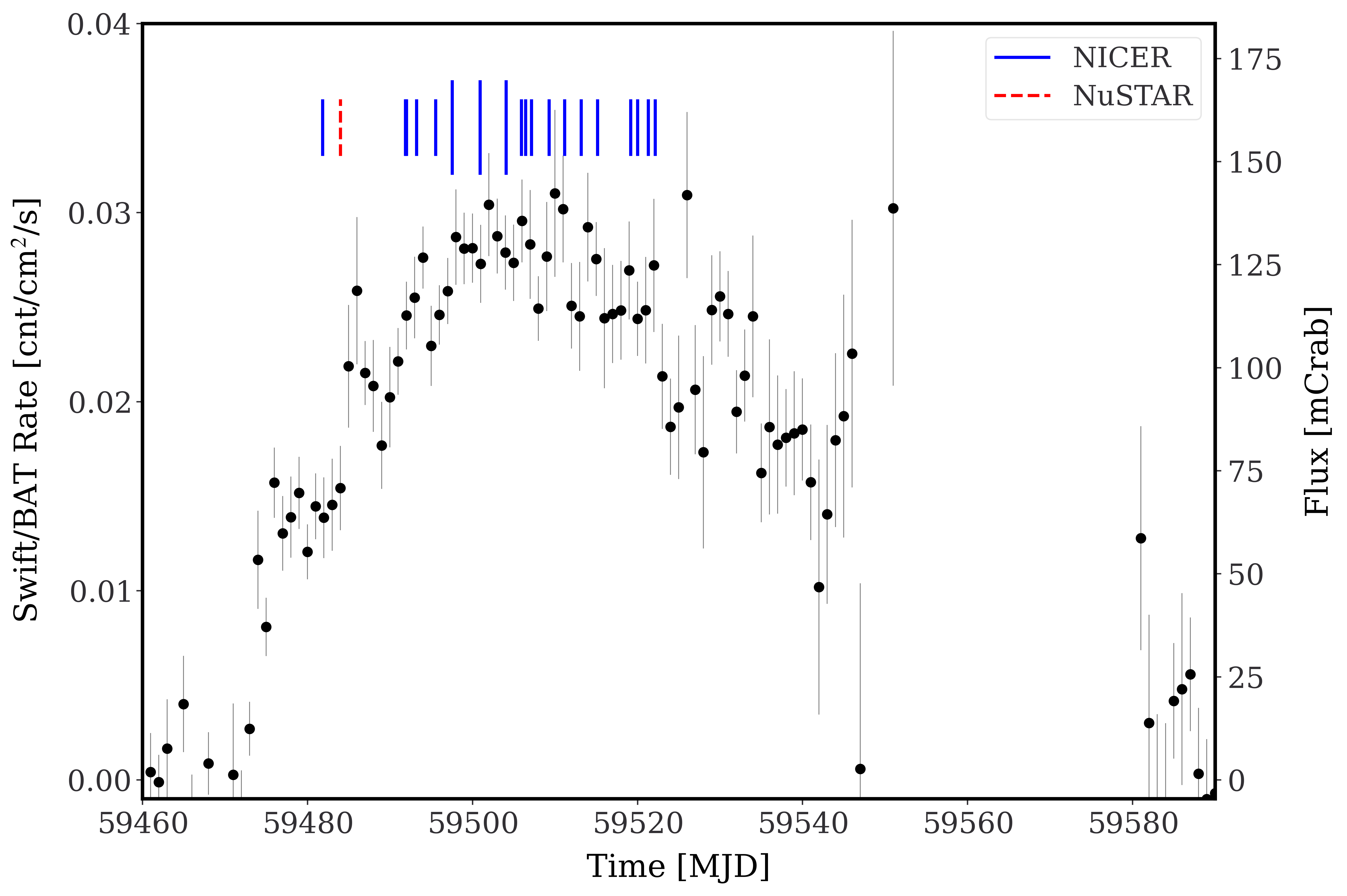}
\caption{\textsl{Swift}/BAT (15-50 keV) daily average light curve of \src during the outburst in 2021 (black dots with gray error bars). Start times of each pointed observation are marked by vertical colored lines as detailed in the legend. 
\textsl{NICER} \mbox{ObsIDs} showing evidence of a secondary peak in the pulse profile (see text) are marked with a longer line.
\label{fig:outburst}}
\end{figure}

\section{Data reduction}\label{sec:data_reduction}

A log of all used observations is shown in Table~\ref{table:log}, while a \textit{Swift}/BAT light curve of the outburst is shown in Fig.~\ref{fig:outburst}.
The outburst started around September 15 2021 (MJD ${\sim}59470$).
\textit{Swift}/BAT data present several observational gaps during the decaying phase, but according to continuous monitoring by Fermi/GBM\footnote{\url{https://gammaray.nsstc.nasa.gov/gbm/science/pulsars/lightcurves/groj1750.html}.}, the source finalized its outburst decay only in January 2022 (MJD ${\sim}59590$).

All data were reduced using instrument-specific pipelines provided by \texttt{HEASOFT} v6.29c.
Spectral data were analyzed using \texttt{XSPEC} v12.12.0 \citep{Arnaud96}.

\begin{table}[!t]
\caption{Log of the source observations used in this work.}
\label{table:log}
\vspace{-0.2cm}
\begin{tabular}{l c c c}
 \hline\\[-2ex]
 & ObsID & MJD & Exposure \\
  & & (Start) & [ks] \\[0.5ex]
  \hline\\[-2ex]
 NuSTAR & 90701331002 & 59484.2 & 29.9 \\[0.5ex] 
 NICER & 4202350101 & 59481.8& 1.7 \\
    & 4202350104 & 59491.8& 1.2 \\
    & 4202350105 & 59491.9& 0.6\\
    & 4202350106 &59493.2 & 2.3\\
    & 4202350107 & 59495.2& 1.7 \\
    & 4202350108 & 59497.5& 1.1\\
    & 4202350109 & 59500.5& 0.1\\
    & 4202350110 & 59504.9& 0.1\\
    & 4202350111 & 59505.1& 0.9\\
    & 4202350112 & 59506.9& 0.9\\
    & 4202350113 &59507.4 & 0.9\\
    & 4202350114 & 59509.1& 1.6\\
    & 4202350115 & 59511.3& 1.6\\
    & 4202350116 &59513.2 & 2.1\\
    & 4202350117 & 59515.2& 1.4\\
    & 4202350118 &59519.2 & 1.8\\
    & 4202350119 & 59520.0& 1.3\\
    & 4202350120 & 59521.3& 0.9\\
    & 4202350121 & 59522.1& 1.9\\[0.5ex] 
 \hline
\end{tabular}
\end{table}

\subsection{NuSTAR}
\textit{NuSTAR} \citep{Harrison13} was launched in 2012 and is currently the only X-ray mission with a telescope able to focus hard X-rays up to 79 keV.
\textit{NuSTAR} consists of two identical co-aligned telescopes that focus X-ray photons onto two independent Focal Plane Modules, FPMA and FPMB.
At the focus of each telescope module are four ($2\times2$) solid-state cadmium zinc telluride (CdZnTe) imaging detectors.
These provide wide-band (3--79\,keV) energy coverage with a FWHM of $18\arcsec$ and a spectral resolution of 400\,eV at 10\,keV, for a field of view of 10 arcmin at 10 keV. 

\textit{NuSTAR} observed \src on September 27, 2021 (ObsID 90701331002, MJD 59484.2) following a Director's Discretionary Time (DDT) request.
The filtered total exposure time was about 30 ks.
\textit{NuSTAR} data were reduced with \texttt{NUSTARDAS} v2.1.1 and using the \textit{CALDB} 20211020 \citep{Madsen21}.
Cleaned events were obtained following the standard \textit{NuSTAR} guidelines.
Source spectra were extracted through the \texttt{NUPRODUCTS} routine.
The source extraction region was a $65\arcsec$ radius circular region centered on the \textit{Chandra} source position.
Due to the position angle (PA) observational constraints and the vicinity of other bright sources, the \textit{NuSTAR} field of view for both modules is contaminated by stray light and ghost rays.
A careful background extraction was therefore applied, in concert with the instrument team.
Given that \textit{NuSTAR} internal background varies from detector to detector, the background should be extracted from the same detector as the source when possible \citep[]{Madsen15}.
We therefore extracted the background from a source-free and contamination-free region of comparable radius to that of the source extraction region on the same detector (Det $0$) and compared it to the background spectrum of other detectors.
This allowed us to verify that the background spectrum was free from artificial, detector-dependent features. 
Moreover, a spectrum of the contaminating flux was extracted and analyzed.
Compared to the source spectrum, the contaminating spectrum was found to be about 300 times fainter, free from prominent features, and dominated by the background counts above 20 keV.
We therefore concluded that our \textit{NuSTAR} source spectrum can be safely considered free from contamination.
The energy band of \textit{NuSTAR} spectral data was limited to the \mbox{$4-60\,$}keV energy range, above which the background counts dominate the spectrum, and to prevent calibration uncertainties in the $3-4\,$keV energy band.

\renewcommand{\arraystretch}{1.2}
\begin{table*}[!t]
\caption{Best-fit results of GRO J1750-27 joint analysis for the joint \textit{NuSTAR} and \textit{NICER} (ObsID 4202350104) spectral data with different best-fit solution for the Model I (Deep and Smooth) \texttt{tbabs*(bbodyrad+highecut*pow+gauss$_{K\alpha}$[+gauss$_{broad}$])*gabs}, Model II \texttt{tbabs*(bbodyrad+cutoffpl+gauss$_{K\alpha}$)*gabs}, and Model III \texttt{tbabs*(bbodyrad+bwcyc+gauss$_{K\alpha}$)*gabs} (see text). All reported errors are at $1\sigma\,$c.l. and based on the MCMC chain values.} \label{table:spectral_all}
\centering
\vspace{-0.2cm}
\begin{tabular}{lccccc}
\hline\\[-3ex]
 & \multicolumn{2}{@{}c}{Model I} & Model II & Model III\\
  & Deep & Smooth & Cutoffpl & BWcyc \\[.05ex]
   \cline{2-3}\\[-2.5ex]
C$_{\rm FPMA}$ (fixed) &  1 & 1 & 1 & 1\\
C$_{\rm FPMB}$ & $1.018^{+0.002}_{-0.002}$ & $1.018^{+0.002}_{-0.002}$ & $1.018^{+0.008}_{-0.002}$ & $1.018^{+0.002}_{-0.003}$\\
C$_{\rm NICER}$ & $1.414^{+0.010}_{-0.007}$ & $1.408^{+0.010}_{-0.007}$ & $1.408^{+0.008}_{-0.008}$ & $1.407^{+0.006}_{-0.008}$\\
N$_{\textrm{H}}$ [$10^{22}\,$cm$^{-2}$] & $4.64^{+0.08}_{-0.05}$ & $8.24^{+0.33}_{-0.09}$ & $4.74^{+0.07}_{-0.06}$ & $6.2^{+0.1}_{-0.2}$\\
kT$_{\rm BB}\,$[keV] & $1.27^{+0.02}_{-0.03}$ & $0.137^{+0.004}_{-0.006}$ & $1.29^{+0.02}_{-0.02}$ & $0.134^{+0.003}_{-0.004}$\\
norm$_{\rm BB}$ & $9.7^{+0.5}_{-0.5}$ & $(6.4^{+5.5}_{-1.2})\,{\times10^6}$ & $9.8^{+0.4}_{-0.4}$ & $(4.1^{+0.7}_{-0.5})\,{\times10^5}$\\
E$_{\rm K\alpha}\,$[keV] & $6.47^{+0.02}_{-0.02}$ & $6.45^{+0.02}_{-0.02}$ & $6.47^{+0.02}_{-0.02}$ & $6.46^{+0.01}_{-0.01}$\\
$\sigma_{\rm K\alpha}\,$[keV] & $0.21^{+0.02}_{-0.02}$ & $0.18^{+0.03}_{-0.02}$ & $0.21^{+0.02}_{-0.02}$ & $0.18^{+0.01}_{-0.02}$\\
norm$_{\rm K\alpha}$ [$10^{-4}$ ph/cm$^2$/s] & $6.3^{+0.5}_{-0.5}$ & $5.3^{+0.6}_{-0.4}$ & $6.1^{+0.4}_{-0.4}$ & $5.2^{+0.1}_{-0.1}$\\
$\Gamma$ & $-1.2^{+0.1}_{-0.1}$ & $1.43^{+0.09}_{-0.01}$  & $-1.2^{+0.1}_{-0.1}$ & --\\
norm$_{\Gamma}^*$ & $(1.2^{+2.9}_{-1.9})\,{\times10^{-3}}$ & $(8.1^{+0.1}_{-0.2})\,{\times10^{-2}}$ & $(1.8^{+0.4}_{-0.3})\,{\times10^{-3}}$ & --\\
HighEcut [keV] & -- & -- & $6.1^{+0.3}_{-0.3}$ & --\\
cutoffE [keV] & $3.2^{+0.5}_{-0.4}$ & $19.1^{+0.8}_{-1.9}$ & -- & --\\
foldE & $6.6^{+0.3}_{-0.1}$ & $13.6^{+2.5}_{-0.5}$ & -- & --\\
E$_{\rm Gauss}\,$[keV] & -- & $5.4^{+0.6}_{-0.9}$ & -- & --\\
$\sigma_{\rm Gauss}\,$[keV] & -- & $10.49^{+1.36}_{-0.05}$ & -- & --\\
norm$_{\rm Gauss}$ [ph/cm$^2$/s] & -- & $0.106^{+0.018}_{-0.004}$ & -- & --\\
$\xi$  & -- & -- & -- & $1.35^{+0.03}_{-0.04}$\\
$\delta$  &  -- & -- & -- & $1.9^{+0.1}_{-0.1}$ \\
B [$10^{12}\,$G]  & --  & -- & -- &$4.62^\star$\\
$\dot{M}$ [$10^{17}\,$g/s]  & -- & -- & -- & $6.74^{+0.11}_{-0.08}$ \\
T$_e$ [keV]  & -- & -- & -- & $4.19^{+0.04}_{-0.05}$ \\
r$_0$ [m]  & -- & -- & --& $87.6^{+5.2}_{-8.0}$\\
d [kpc]  &  -- & -- & --& 14.0 (fixed)\\
norm$_{bwcyc}$&  -- & -- & --& $(7.7^{+0.6}_{-0.9})\,{\times10^{-2}}$\\
E$_{\rm cyc}\,$[keV] & $44.0^{+0.8}_{-0.3}$ & $43.3^{+1.0}_{-0.5}$  & $44.0^{+0.6}_{-0.6}$ & $43.6^{+0.3}_{-0.2}$\\
$\sigma_{\rm cyc}\,$[keV] & $10.0^{+0.6}_{-0.2}$ & $6.3^{+1.4}_{-0.5}$ & $9.6^{+0.6}_{-0.5}$ & $10.5^{+0.9}_{-0.2}$\\
Strength$_{\rm cyc}^\ddagger$ ($\tau_{cyc}^\ddagger$ [keV]) & $50.6^{+1.3}_{-0.7}\, (2.02^{+0.09}_{-0.14})$ & $19.8^{+6.5}_{-3.1}\, (1.6^{+0.4}_{-0.2})$ & $46.2^{+4.9}_{-5.6}\, (1.9^{+0.3}_{-0.3})$ & $46.7^{+4.2}_{-2.7}\, (1.8^{+0.2}_{-0.2})$\\
Flux$^\dagger$ (1-60 keV)& $2.336^{+0.005}_{-0.004}$ & $2.52^{+0.01}_{-0.01}$ & $2.346^{+0.005}_{-0.005}$ & $2.427^{+0.007}_{-0.006}$ \\
$\chi^2$/d.o.f. & $3156/3395$ & $3308/3392$ & $3364/3396$ & $3435/3393$ \\
Goodness-of-fit & $44\%$ & $45\%$ & $40\%$ & $45\%$\\
Null-hypothesis P & $70\%$ & $85\%$ & $68\%$ & $28\%$\\
\hline
\end{tabular}
\vspace{-0.2cm}
\tablefoot{
$^*$In units of photons/keV/cm$^2$/s at 1 keV.\quad$^\star$Linked to the E$_{\rm cyc}$ parameter (see text).\quad$^\ddagger$Line depth (optical depth) as defined in \texttt{XSPEC}.\quad$^\dagger$Unabsorbed flux calculated for the entire model (in units of $10^{-9}\,$erg\,cm$^{-2}\,$s$^{-1}$), obtained using the \texttt{cflux} model from \texttt{XSPEC} as resulting from FPMA.}
\end{table*}


\subsection{NICER}

\textit{NICER} \citep{Gendreau2017} is an X-ray telescope deployed on the International Space Station (ISS) in 2017 June. \textit{NICER} X-ray Timing Instrument (XTI) has 56 aligned focal plane modules FPMs (52 currently operational), each made up of an X-ray concentrator optic associated with a silicon drift detector.
The peak collecting area of all FPMs combined is $1900\,$cm$^2$ at 1.5 keV, with a field of view of 30 arcmin$^2$. \textit{NICER} is capable of fast-timing observations in the $0.2-12.0$ keV band, with timing accuracy of time-tagged photons to better than 100 ns \citep{Prigozhin2016, LaMarr2016, Gendreau2017, Okajima2016}. The spectral resolution is about 140 eV at 6 keV.

Upon Target of Opportunity (ToO) request, \textit{NICER} monitored GRO J1750-27 also in coordination with \textit{NuSTAR}.
However, simultaneous \textit{NICER} observations were prevented by ISS visibility constraints.
\textit{NICER} started observing GRO J1750-27 on September 24, 2021 (ObsID 4202350101, MJD 59481.8), and continued to monitor the source with variable cadence until November 4, 2021 (ObsID 4202350121).
The total exposure time was about 24 ks after data cleaning.
Hereinafter, \textit{NICER} ObsIDs are identified by their last two digits, that is 42023501XX.
We also verified that \textit{NICER} data were not contaminated by other sources in the relatively large field of view. In fact, the field of view includes the supergiant fast X-ray transient IGR J17503-2636 \citep[]{Ferrigno2019}.
However, the same sky region was also monitored by \textit{INTEGRAL}, and IGR J17503-2636 was not X-ray active during our \textit{NICER} observations.

NICER data were processed with HEASoft version 6.29c and the \textit{NICER} Data Analysis Software (\texttt{nicerdas}) version 8 (\texttt{2021-08-31\_V008c}) with Calibration Database (CALDB) version \texttt{xti20210720}, adopting standard calibration and screening criteria from the \texttt{nicerl2} tool.
The adopted version of \texttt{nicerdas} also produces proper response files (RMF and ARF) for each ObsID.
The background spectrum was obtained using the \texttt{nibackgen3C50 v7} tool\footnote{\url{https://heasarc.gsfc.nasa.gov/docs/nicer/tools/nicer_bkg_est_tools.html}} \citep[]{Remillard21}.
Hot detectors and additional off detectors in a few ObsIDs were also identified and manually excluded during the data-reduction pipeline.
ObsIDs 02 and 03 were not considered here due to their null exposure time.
As advised by the instrument team, a systematic error of 1\% has been applied for all \textit{NICER} spectra.
The energy band of \textit{NICER} spectra was limited to $0.7-10\,$keV, below which background counts dominate the spectrum, and to prevent calibration uncertainties in the $10-12\,$keV energy band.

\renewcommand{\arraystretch}{1.2}
\begin{table*}[!t]
\caption{Best-fit results of GRO J1750-27 \textit{NICER}-only spectral analysis using a cutoff power-law model \texttt{cutoffpl} with an iron K$\alpha$ line. All reported errors are at $1\sigma\,$c.l., obtained using the \texttt{err} tool from \texttt{XSPEC}.} \label{table:spectral_nicer}
\centering
\vspace{-0.2cm}
\begin{tabular}{lccccccccccc}
\hline\\[-3ex]
ObsID & N$_{\rm H}$ & $\Gamma$ & HighEcut & norm$^*_\Gamma$& Flux$^\dagger$ & $\chi^2/\rm d.o.f.$ & E$_{\rm K\alpha}\,$ & $\sigma_{\rm K\alpha}\,$ & norm$_{\rm K\alpha}^\star$ \\
  & $[10^{22}\,$cm$^{-2}]$ & & [keV] &  & & & [keV] & [keV] &  &  \\[0.05ex]
  \hline\\[-2ex]
4202350101 & $4.8(2)$ & $0.06(12)$ & $10.3_{-1.9}^{+3.0}$ & $1.6(2)$ & $0.593(9)$ & $1078/944$ & -- & -- & -- \\
4202350104 & $5.2(1)$ & $0.01(12)$ & $7.7_{-1.1}^{+1.4}$ & $2.8(2)$ & $1.11(2)$ &$1035/942$ & $6.53(8)$ & $0.16(6)$ & $0.05(2)$\\
4202350105 & $5.4(2)$ & $0.1(2)$ & $8.9_{-1.8}^{+3.2}$ & $3.4(4)$ & $1.14(2)$ &$1097/935$ & $6.59(9)$ & $0.21(8)$ & $0.1(3)$\\
4202350106 & $5.3(1)$ & $0.03(8)$ & $8.5_{-0.9}^{+1.1}$ & $3.4(2)$ & $1.24(1)$ & $1138/942$ & $6.56(4)$ & $0.20(4)$ & $0.11(2)$\\
4202350107 & $5.3(1)$ & $0.03(8)$ & $8.0_{-0.9}^{+1.2}$ & $3.8(2)$ & $1.31(1)$ &$1066/942$ & $6.53(4)$ & $0.14(4)$ & $0.08(2)$\\
4202350108 & $5.3(1)$ & $-0.05(10)$ & $7.4_{-0.9}^{+1.2}$ & $3.7(2)$ & $1.39(1)$ &$1070/942$ & $6.61(8)$ & $0.20(7)$ & $0.09(2)$\\
4202350109 & $6.1(4)$ & $0.6(2)$ & $33 (unc.)$ & $6.3(6)$ & $1.55(7)$ &$1006/887$ & $6.33(3)$ & $0.1 (unc.)$ & $0.04(3)$\\
4202350110 & $5.7(5)$ & $0.3(3)$ & $14_{-5}^{+15}$ & $4.7(9)$ & $1.49(7)$ &$1136/877$ & $6.47(7)$ & $0.4(1)$ & $0.3(1)$\\
4202350111 & $5.5(2)$ & $-0.01(12)$ & $7.2_{-0.9}^{+1.3}$ & $4.7(4)$ & $1.58(2)$ &$1036/919$ & $6.56(9)$ & $0.19(8)$ & $0.09(3)$\\
4202350112 & $5.2(2)$ & $-0.17(12)$ & $6.1_{-0.7}^{+0.9}$ & $3.8(3)$ & $1.45(2)$ &$992/942$ & $6.6(1)$ & $0.3(1)$ & $0.098(4)$\\
4202350113 & $5.3(2)$ & $-0.06(12)$ & $7.2_{-0.9}^{+1.3}$ & $3.8(3)$ & $1.40(2)$ &$1133/941$ & $6.4(1)$ & $0.01(1)$ & $0.04(1)$\\
4202350114 & $5.4(1)$ & $0.08(9)$ &
$8.7_{-1.1}^{+1.4}$ & $4.3(3)$ & $1.46(2)$ &$1093/942$ & $6.50(4)$ & $0.17(4)$ & $0.10(2)$\\
4202350115 & $5.0(1)$ & $-0.17(9)$ & $6.2_{-0.6}^{+0.7}$ & $3.9(3)$ & $1.48(1)$ & $1045/942$ & $6.52(6)$ & $0.23(6)$ & $0.12(3)$ \\
4202350116 & $5.3(1)$ & $-0.22(9)$ & $5.4_{-0.4}^{+0.5}$ & $3.3(3)$ & $1.23(1)$ & $1076/942$ & $6.62(6)$ &  $0.23(6)$ & $0.09(2)$ \\
4202350117 & $5.4(1)$ & $0.08(9)$ & $8.8_{-1.2}^{+1.7}$ & $4.2(3)$ & $1.44(1)$ &$987/942$ & $6.51(5)$ & $0.15(5)$ & $0.09(2)$\\
4202350118 & $5.5(1)$ & $0.13(9)$ & $9.4_{-1.3}^{+1.8}$ & $3.7(3)$ & $1.22(2)$ &$1027/942$ & $6.45(6)$ & $0.16(8)$ & $0.08(2)$\\
4202350119 & $5.4(1)$ & $0.06(11)$ & $8.3_{-1.2}^{+1.6}$ & $3.5(3)$ & $1.18(2)$ &$926/932$ & $6.56(9)$ & $0.17(9)$ & $0.05(2)$\\
4202350120 & $5.7(2)$ & $0.17(14)$ & $9.1_{-1.7}^{+2.7}$ & $3.8(4)$ & $1.13(2)$ &$1055/892$ & $6.52(6)$ & $0.02(9)$ & $0.03(2)$\\
4202350121 & $5.7(1)$ & $0.40(9)$ & $21_{-6}^{+12}$ & $3.9(3)$ & $1.18(1)$ &$945/932$ & $6.64(8)$ & $0.29(9)$ & $0.10(2)$\\[1ex]
\hline
\end{tabular}
\vspace{-0.2cm}
\tablefoot{
$^*$In units of $10^{-2}\,$photons/keV/cm$^2$/s at 1 keV.\quad$^\dagger$Unabsorbed flux calculated for the entire model in the $0.5-10\,$keV energy band and reported in units of $10^{-9}\,$erg\,cm$^{-2}\,$s$^{-1}$.\quad$^\star$In units of $10^{-2}\,$photons/cm$^2$/s. \quad$^{unc.}$ Unconstrained value.}
\end{table*}

\section{Data analysis and results}

\subsection{Spectral analysis}
\subsubsection{Spectral setup}\label{subsec:spectral}

Both background and source spectra were rebinned to have at least one count per bin (similarly to, e.g., \citealt[]{Snios2020}) in order to use the C-statistic \citep[]{Cash79} with Poissonian background (W-stat) as a fit statistic.
Such rebinning addresses possible biases when using W-stat\footnote{\url{https://heasarc.gsfc.nasa.gov/xanadu/xspec/manual/XSappendixStatistics.html}}.
Given the non-Poissonian nature of NICER background \citep[]{Remillard21}, the \texttt{pgstat} statistic was used for NICER data.
On the other hand, the $\chi^2$ was employed as a test statistic.
For all tested models, the photoelectric absorption component was set according to \citet[\texttt{tbabs} in \texttt{XSPEC}]{Wilms00} to account for photoelectric absorption by neutral interstellar matter (or column density N$_{\rm H}$), and we assumed model-relative (\texttt{wilm}) solar abundances.
The Galactic N$_{\rm H}$ in the direction of the source is about \mbox{$1.1\times10^{22}\,$cm$^{-2}$} \citep{HI4PI2016}.

We performed spectral analysis both for the joint \textit{NICER} and \textit{NuSTAR} spectra (using data from the neighbor \textit{NICER} ObsID 04, see Sect.\,\ref{subsubsec:joint}), and for the single \textit{NICER} observations (see Sect.\,\ref{subsubsec:singles}).
In the following, errors are reported at $1\sigma\,$c.l. in tables and figures. 
Errors for the joint \textit{NICER} and \textit{NuSTAR} spectral analysis are calculated through MCMC simulations using the Goodman-Weare algorithm of length $2\times10^5$ with 20 walkers and $10^4$ burn-in steps.
For the joint fit, the goodness-of-fit parameter is also reported as resulting from the \texttt{goodness} tool from XSPEC (with \texttt{fit} and \texttt{nosim} options), indicating the percentage of simulated spectra with a test statistic better than the one obtained from real data.
In those cases, for each relevant model, we also report the null-hypothesis probability (NHP) as resulting from the \texttt{XSPEC} fitting procedure, which represents the probability of the data being drawn from the best-fit model.
For the single \textit{NICER} ObsIDs spectral analysis, errors are calculated through the \texttt{err} tool in \texttt{XSPEC} without any chains loaded, as 
this method is less computational expensive (compared to MCMC simulations) and, at the same time, reliable for steady spectra such as those analyzed here (which do not show any local minima in their best-fit solution).
Our spectral analysis results are reported in Sect.~\ref{subsubsec:joint} and  Table~\ref{table:spectral_all} for the joint \textit{NICER} and \textit{NuSTAR} analysis and in Sect.~\ref{subsubsec:singles}  and Table~\ref{table:spectral_nicer} for \textit{NICER}-only analysis.

\subsubsection{Joint \textit{NICER} and \textit{NuSTAR} spectral analysis}\label{subsubsec:joint}

\citet[]{MalacariaAtel2022} reported the GRO J1750-27 spectral fitting results for the joint \textit{NICER} and \textit{NuSTAR} data analysis. The best-fit continuum spectrum presented by those authors consists of an absorbed power law with high-energy cutoff (\texttt{highecut*powerlaw}) modified by an ad hoc broad Gaussian emission component (\texttt{gauss}) that serves two purposes. First, it smooths the steep high-energy cutoff, thus providing a good fit of data around the cutoff energy.
Second, the spectrum also shows an absorption feature at about 43 keV, modeled with a Gaussian absorption line and interpreted as an electron CRSF.
However, the cyclotron line depth is found to be unusually large (i.e., optical depth $\tau>2$) if the ad hoc broad component is not added to the model spectrum. The addition of the broad Gaussian emission component therefore mitigates the otherwise exceptionally deep cyclotron line.

\begin{figure}[!t]
\includegraphics[width=.47\textwidth]{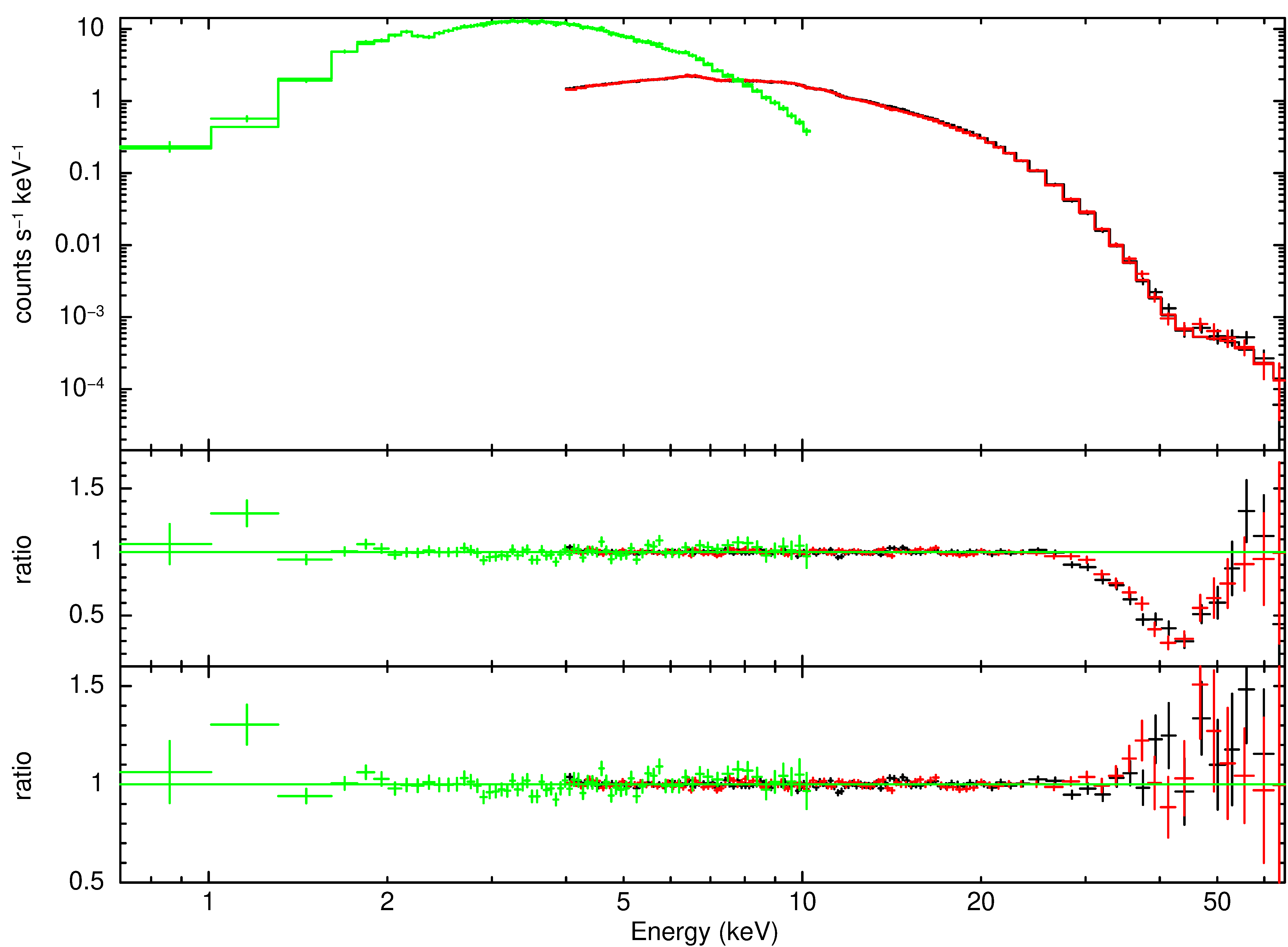}
\caption{\textit{NICER} (green) and \textit{NuSTAR} (black and red) combined spectrum of \src. Top panel: Data (crosses) and best-fit model (continuous lines). Middle: Ratio residuals (data divided by model) using Model I Smooth (see Table\,\ref{table:spectral_all}) but without the inclusion of a cyclotron line. Bottom: Residuals from the best-fit model including a Gaussian absorption line near 43 keV to account for the cyclotron resonant scattering feature. Data were rebinned for plotting purposes.\label{fig:spectrum}}
\end{figure}

The best-fit spectrum also includes an iron K$\alpha$ line at 6.4 keV.
Finally, positive residuals remaining around 1 keV were modeled with a blackbody component, which returned a lower $\chi^2$ compared to a partial covering component.
The entire model (see Fig.~\ref{fig:spectrum} and ``Model I Smooth'' in Table~\ref{table:spectral_all}) fits the data well, but one disadvantage is that it needs the broad Gaussian emission component, which mimics a high-energy bump but has no physical interpretation.
For completeness, Table~\ref{table:spectral_all} also reports the results of the best-fit model without the broad Gaussian emission component, which results in a deeper CRSF (``Model I Deep'').

\begin{figure}[!t]
\includegraphics[width=.47\textwidth]{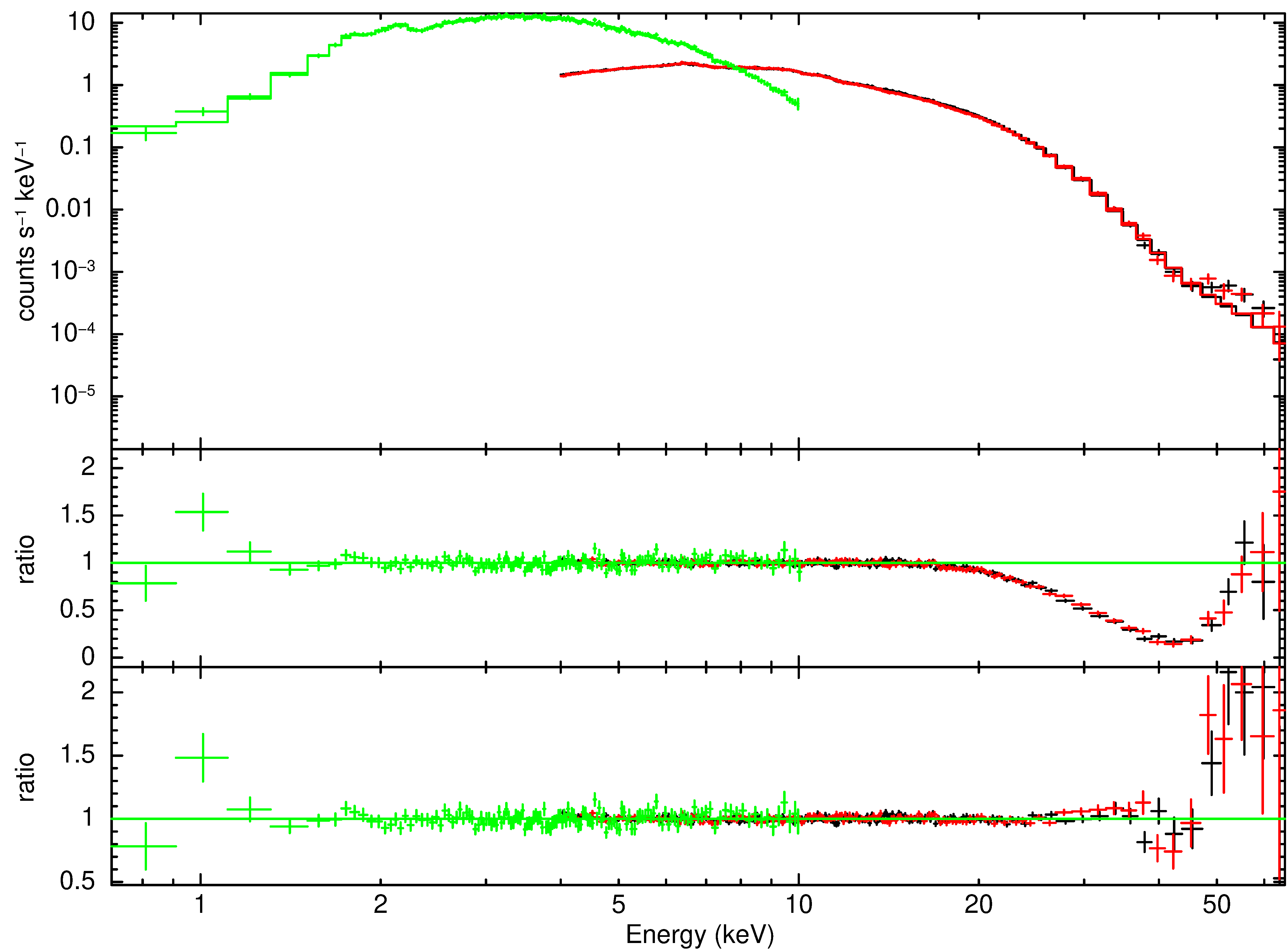}
\caption{Same as Fig.~\ref{fig:spectrum}, but for the \texttt{bwcyc} model (see Table~\ref{table:spectral_all}). \label{fig:spectrum2}}
\end{figure}

\begin{figure}[!t]
\includegraphics[width=.47\textwidth]{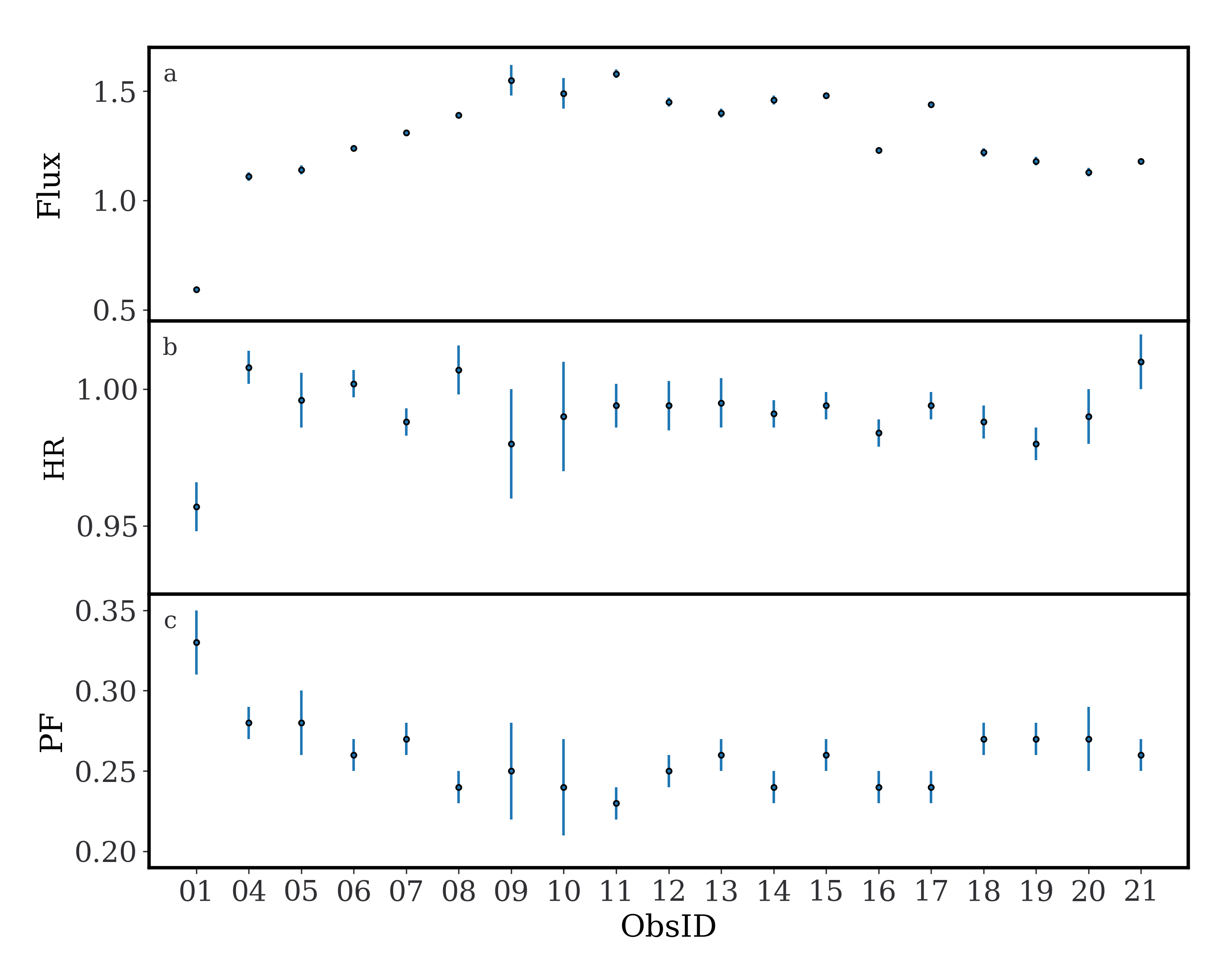}
\caption{Evolution of the investigated fundamental quantities for \src
as observed by \textit{NICER} in each ObsID (x-axis) of the outburst monitored in 2021.
\textit{Top panel (a): }Flux in the $1-10\,$keV band. \textit{Middle panel (b): }Hardness ratio [${4-10}\,$keV/${1-4}\,$keV]. \textit{Bottom panel (c): }Pulsed fraction.
\label{fig:total_plots}}
\end{figure}
Building upon those results, we also searched for alternative and possibly more physically consistent spectral models (i.e., not requiring the ad hoc broad Gaussian emission component) that also return a more commonly observed cyclotron line depth.
Our analysis takes advantage of the broader energy coverage including \textit{NICER} data with respect to \citet[]{Sharma22} and \citet{Devaraj22}, whose analysis is limited to the \textit{NuSTAR} energy band. This is significant given the peculiar steepness of the spectral continuum and the complex cyclotron line profile, both of which can be better constrained thanks to the lower-energy continuum data.
The following models were therefore tested: an absorbed Comptonization model of soft photons in a hot plasma (\texttt{CompTT} in \texttt{XSPEC}, \citealt{Titarchuk94}), \texttt{cutoffpl}, \texttt{highecut*powerlaw}, a power-law modified by a Fermi-Dirac cutoff \texttt{powerlaw*FDcut} \citep[]{Tanaka86}, and the Negative and Positive power laws with a common EXponential cutoff \texttt{NPEX} \citep[]{Mihara98}. Of these, the best-fit model showing the highest NHP is the cutoff power-law model (see Table\,\ref{table:spectral_all}).
The spectral model \texttt{highecut*powerlaw} was also employed in other works to fit the continuum of \src as observed by \textit{NuSTAR}, although modified by the inclusion of different components, such as a nonharmonic cyclotron line \citep[]{Sharma22} or a smoothing absorption feature centered at the cutoff energy \citep[]{Devaraj22}.
However, we also separately tested the nonharmonic cyclotron line and the smoothing absorption feature around the cutoff energy but, as also reported by the respective authors, we noticed that their inclusion only slightly reduces the fundamental line depth. Therefore, we opted to focus on a different model that does not include additional absorption components, and to report the results of the cutoff power-law model in Table\,\ref{table:spectral_all}.

Moreover, two physical models for spectral formation based on bulk and thermal Comptonization in accreting XRPs were also tested, namely the \texttt{bwcyc} model \citep[]{Becker+Wolff07, Ferrigno07} and the \texttt{compmag} model \citep[]{Farinelli12}.
Of these, the \texttt{bwcyc} model returned a $\chi^2$ value much lower than the \texttt{compmag} model (the difference in test statistic between the two models was $\Delta\chi^2\sim400$), and the former was therefore employed in our analysis (see Fig.~\ref{fig:spectrum2} and Table~\ref{table:spectral_all}). Following the official usage guidelines, we kept the following model parameters frozen during the fit: NS radius R$_{\rm NS}$ and mass M$_{\rm NS}$, distance to the source $D$, and magnetic field strength $B$. For those parameters we chose the following values: R$_{\rm NS}=12\,$km, M$_{\rm NS}{=}1.4\,\rm M_{\odot}$, $\rm D{=}14\,$kpc, ${\rm  B{=}E_{cyc}^{best-fit}} ((1+z)/11.6) {\times}10^{12}\,$G (where $\rm E_{\rm cyc}^{\rm best-fit}$ is the best-fit centroid energy of the cyclotron line at about 44 keV, and $z\approx0.24$ is the gravitational redshift).
Among the other model parameters are the so-called similarity parameters $\xi$ and $\delta$. These describe, respectively, the importance of the photons escape through the accretion column and the relative importance of bulk and thermal Comptonization:
\begin{equation}
    \quad\quad\quad\quad \xi = \frac{\pi r_0 m_p c}{\dot{M}\sqrt{\sigma_\parallel \sigma_\perp}}, \quad\quad\quad \delta = 4\frac{y_{bulk}}{y_{therm}} \propto \frac{\sigma_\parallel}{\bar{\sigma}}. 
\end{equation}

The other model parameters are: the accretion column radius $r_0$, the mass accretion rate $\dot{M}$, and the electron temperature of the Comptonizing electrons $T_e$, while $m_p$ is the proton mass, $y_{bulk}$ and $y_{therm}$ are the Compton $y-$parameters as defined in \citet[]{Becker+Wolff07}, and $\bar{\sigma}$ and $\sigma_\parallel$ are the electron scattering cross-sections angle-averaged and parallel to the magnetic field, respectively.
The above-mentioned frozen set of values represents the best fit when compared to other configurations of the same model where the distance was frozen at 16 or 18 kpc.

\subsubsection{NICER ObsIDs spectral analysis}\label{subsubsec:singles}

\begin{figure}[!t]
\includegraphics[width=.47\textwidth]{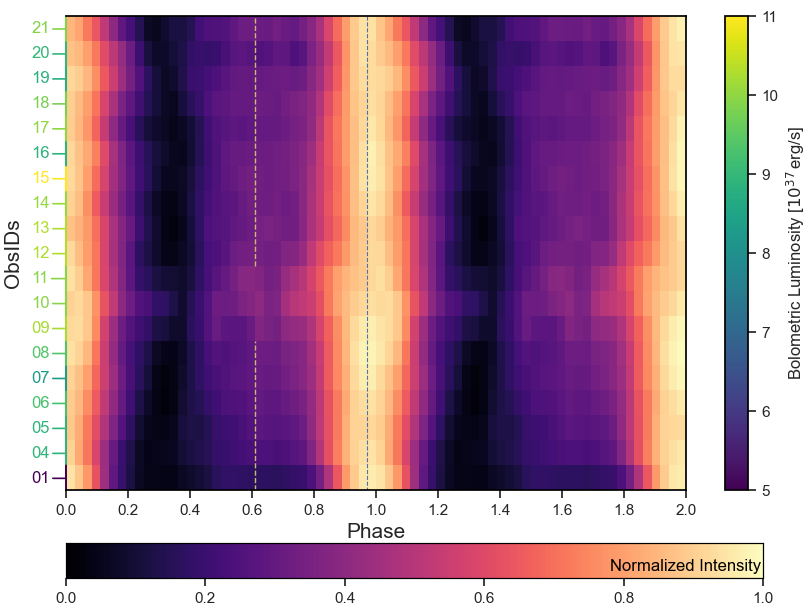}
\caption{Heat map of \textit{NICER} pulse profiles from GRO~J1750-27 as monitored during the 2021 outburst. The black vertical dashed line at spin phase ${\sim}0.95$ marks the pulse profile peak for ObsID 01. The bottom horizontal bar shows the color-coded normalized intensity. The color-coded ObsIDs reported on the left y-axis show the bolometric luminosity (as derived by \textit{Swift}/BAT count rates, see text) at each ObsID as illustrated by the color bar to the right. The vertical yellow dashed line at~spin phase ${\sim}0.6$ confines the ObsIDs where the secondary peak appears more prominently. Pulse profiles are plotted twice for clarity and bins have been smoothed with a Gaussian filtering for visibility purpose.
\label{fig:heatmap}}
\end{figure}

\textit{NICER} spectral data from each ObsID have also been modeled to investigate spectral variations throughout the observed outburst stages.
Due to the restricted energy band, only a relatively simple model was tested, namely an absorbed cutoff power-law model with the addition of a Gaussian emission line for the Fe K$\alpha$ around 6.4 keV. Results are reported in Table\,\ref{table:spectral_nicer}.
The model can fit the data at all ObsIDs, returning best-fit spectral continuum parameters that are roughly consistent throughout all observations.

The iron K$\alpha$ line in GRO J1750-27 as observed by NICER has an energy of about $6.3-6.6\,$keV. 
This is consistent with emission of fluorescent lines from neutral or weakly ionized iron.
The only observation that did not require an iron line component (i.e., normalization consistent with zero) is the ObsID 01, at the lowest flux observed by \textit{NICER}.
The nondetection of the iron line in ObsID 01 was also tested through Monte Carlo simulations. To this aim, the \texttt{XSPEC simftest} routine was employed, which allows the user to simulate a given number of spectra based on the actual data and test the resulting $\Delta\chi^2$ between each instance of the fit with and without the model component to be tested (the Gaussian line in our case). A total of $10^4$ simulations were carried out. The obtained $\Delta\chi^2$ from fitting the actual data with and without the iron line is about 6, but about $10\%$ of the simulations show a higher $\Delta\chi^2$, thus suggesting that the observed improvement is likely due to statistical fluctuations (significance $<2\sigma$).

To further investigate the spectral behavior, we explored the spectral evolution in a model-independent way, that is by means of the hardness ratio (HR).
We define the HR as the count-rate ratio in two different energy bands, i.e., HR$\,=4-10$ keV / $1-4$ keV.
Fig.~\ref{fig:total_plots}(a and b) shows the evolution of the HR as observed during \textit{NICER} observations. Except for the first ObsID, there is no significant variation of the HR along the outburst.

\begin{figure}[!t]
\includegraphics[width=.47\textwidth]{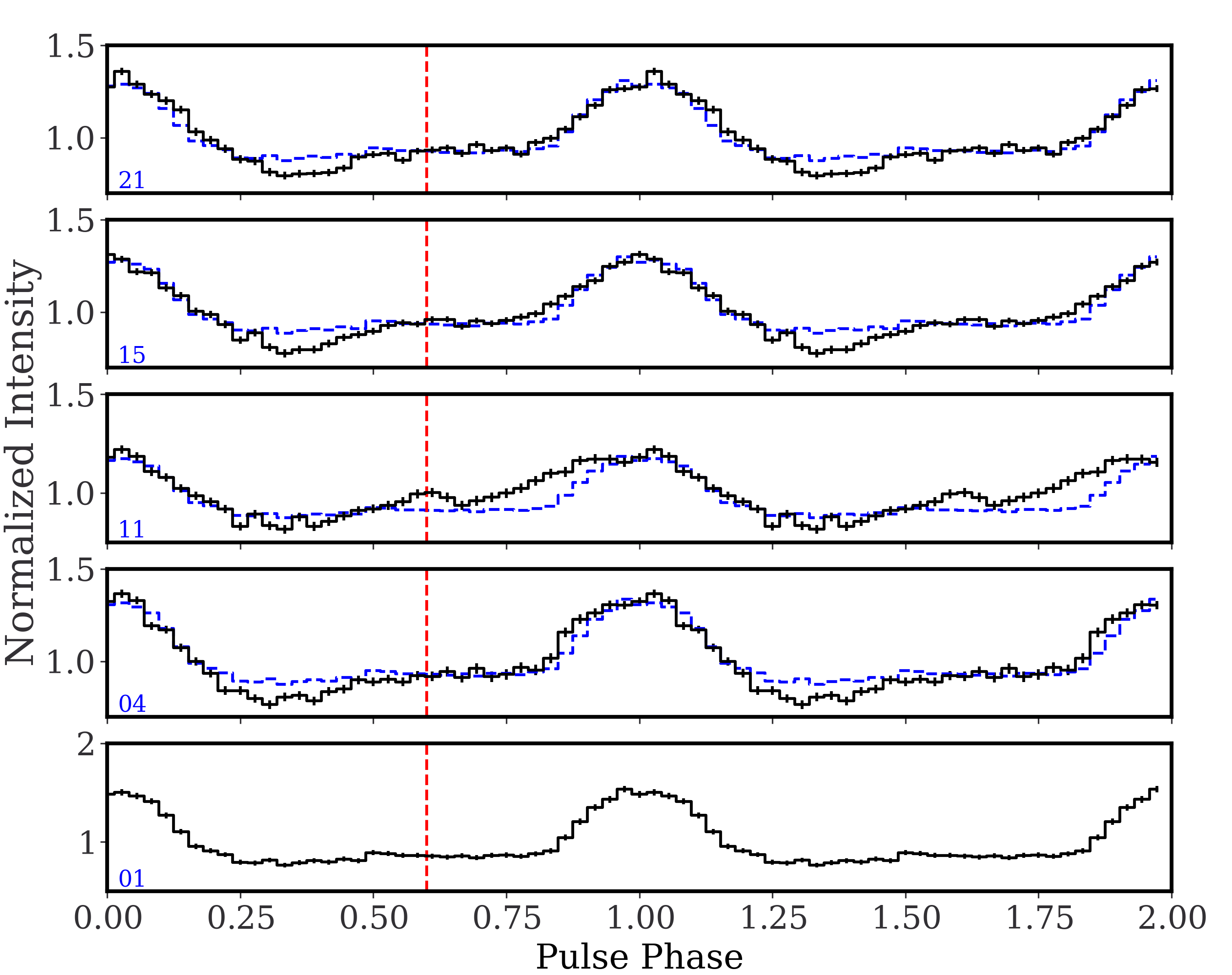}
\caption{Selection of representative \textit{NICER} pulse profiles. The labels on the bottom left corner of each panel represent the correspondent ObsID. The red dashed vertical line marks the secondary peak that is more relevant in ObsID 11 at $\phi=0.6$ (see also Fig.\,\ref{fig:heatmap}). The blue dashed line shows the pulse profile from ObsID 01 rescaled for reference. Pulse profiles are plotted twice for clarity and normalized by the average source intensity in a given ObsID. \label{fig:profiles}}
\end{figure}

\subsection{Timing analysis}\label{subsec:timing}

To analyze the \textit{NICER} pulse profiles at different outburst stages, barycentered events were extracted in the $1-10\,$keV energy band using JPL-DE405 Solar System ephemeris.
We then searched for pulsations for each ObsID using the epoch folding method \citep[]{Leahy+83} as employed by the \texttt{efsearch} tool in \texttt{HEASOFT}.
For reference, a spin period of 4.45150(1) was obtained for ObsID 01 (the pulse period uncertainty was estimated through simulations following the method outlined in \citealt[]{Lutovinov12, Boldin13}).
Events from each ObsID were then folded with their correspondent pulse period to create pulse profiles.
To phase-align the pulse profiles, a cross-correlation was performed between two consecutive ObsID profiles.
The phase-aligned pulse profiles are shown in the heat map in Fig.~\ref{fig:heatmap}.

By comparing the flux obtained by the joint \textit{NICER} and \textit{NuSTAR} spectral analysis ($2.346(5)\times10^{-9}\,$erg\,cm$^{-2}\,$s$^{-1}$, see Table\,\ref{table:spectral_all}, Model II) with the contemporary Swift/BAT count rate (0.015(2) cnt\,cm$^{-2}\,$s$^{-1}$) at MJD 58484, we obtain a conversion factor $C_f=1.5(2)\times10^{-7}\,$erg\,cm$^{-2}$ to estimate the $1-60\,$keV flux from the Swift/BAT count rate, which we use as a proxy for the bolometric flux.
Such an estimate represents a valid approximation if the spectral emission does not change its shape, a condition that is well satisfied here (see Sect.~\ref{subsubsec:singles}).
By applying this conversion factor, and assuming a distance value of 14 kpc (see Sect.~\ref{subsec:spectral}), a corresponding luminosity value was associated to each \textit{Swift}/BAT data point and therefore to each correspondent \textit{NICER} observation. This information was included in the \textit{NICER} pulse-profile heat map in Fig.~\ref{fig:heatmap}.
A selection of representative \textit{NICER} pulse profiles is also shown in Fig.~\ref{fig:profiles}.

Finally, we also investigated the evolution of the pulsed fraction (PF) during the outburst.
For this, the PF was defined as 
($I_{\rm max}-I_{\rm min}$)/($I_{\rm max}+I_{\rm min}$), where $I_{\rm max}$ and $I_{\rm min}$ are the maximum and minimum pulse-profile count rates, respectively.
The observed PF evolution is shown in Fig.~\ref{fig:total_plots}c.

\section{Discussion}

\subsection{The broadband spectrum}

The joint \textit{NuSTAR} and \textit{NICER} spectral analysis of \src allowed us to confirm the discovery of a cyclotron line at about 44 keV.
For cyclotron lines, the centroid energy of the fundamental line is linked to the magnetic field as E$_{cyc}\approx\,B_{12}\, 11.6\,E_{cyc}/(1+z)$ (where B$_{12}$ is the magnetic field strength in units of $10^{12}\,$G).
This corresponds to B$\simeq4.7\times10^{12}\,$G.
This value is in agreement with recent findings from \citep[]{MalacariaAtel2022,Devaraj22,Sharma22}.
The cyclotron line also shows an unusually large optical depth ($\tau\simeq2$, see Table~\ref{table:spectral_all}).
For comparison, some of the deepest fundamental cyclotron lines were observed in A0535+26 \citep[]{Grove95} and V~0332+53 \citep[]{Mowlavi06}, with optical line depths $\tau=1.8$ and $\tau=2.11$, respectively\footnote{The optical depth values are derived here following the definition of the \texttt{gabs} function in \texttt{XSPEC}.}. V~0332+53 also shows some of the deepest harmonic lines \citep[]{Pottschmidt05}, with $\tau=3.3$ for the second harmonic at $\sim74\,$keV. More commonly, the average cyclotron line depth is $\tau=0.5-1$ \citep[]{Staubert19, Malacaria21, Malacaria2022}, although one must also account for the anti-correlation of the line depth with luminosity \citep{Mowlavi06,Tsygankov+10}.
However, it is also worth noting that even super-Eddington XRPs do not show unusually large values of the cyclotron line depth \citep[]{Jaisawal2016, Kong2022}.
Moreover, we note that the line depth in GRO J1750-27 remains large with most spectral models tested here and elsewhere \citep[]{Sharma22, Devaraj22}.
All these aspects support the interpretation that the large line depth has a physical origin and is not merely an artefact of the imperfect spectral modeling.

Several spectral models have been tested for the broadband emission from GRO J1750-27.
For all tested spectral models, the best-fit column absorption value is several times larger than the Galactic absorption in the direction of the source, in agreement with \citet[]{Lutovinov19}.
We found that a cutoff power-law model can fit the data with a negative power-law photon index (see Model I Deep and Model II in Table~\ref{table:spectral_all}).
Although a positive power-law photon index is typically observed in XRPs, a negative photon index has been observed in a few sources \citep[]{Iyer15, Aftab19} where the spectrum is generally harder than the typical XRP spectrum. Here, when the analysis is limited to the narrower \textit{NICER} energy band (see Sect.~\ref{subsubsec:singles} and Table~\ref{table:spectral_nicer}), the obtained photon index is in general much softer than that obtained when the broadband spectrum is considered (i.e., Model I Deep).
However, the negative photon index in Model I Deep is softened when the emission excess around the cutoff energy is modeled with a broad Gaussian emission component (see Model I Smooth in Table~\ref{table:spectral_all}).
This way, the obtained photon index value is in agreement with the more commonly observed positive values.
Such a hard bump is reminiscent of the 10-keV feature observed in many accreting XRPs \citep[]{Coburn+02}.
Even so, the hard bump feature in GRO J1750-27 is broader than the 10-keV width observed elsewhere (see, e.g., \citealt{Klochkov07,Ferrigno09}).
Just as there is no commonly accepted interpretation of the 10-keV feature (that is, its physical origin in unknown), the hard bump observed in GRO J1750 is likely due to the inadequacy of the adopted phenomenological model, but it nonetheless highlights the fact that the observed spectral shape for GRO J1750-27 differs from those commonly observed in accreting XRPs.

We also notice that the blackbody component best-fit parameters return two different solutions according to the employed continuum model.
One solution (see Model I Deep and Model II in Table~\ref{table:spectral_all}), similar to the solution obtained by \citealt{Sharma22}, consists of a blackbody component with $kT\sim 1.3\,$keV and an emitting radius of about 4 km (assuming a distance of 14 kpc).
\citealt{Sharma22} ascribe this component to the NS surface. 
However, as also pointed out by \citealt{Hickox04} for example, it is worth noting that the blackbody emission from the NS surface is typically colder (i.e., $\sim0.1\,$keV) and only extends to the accreting polar cap (i.e., $\sim10^2\,$m, see also below).
Even when hot (e.g., $1.3\,$keV) blackbody components have been ascribed to the NS surface, their emitting radius was found to be consistent with the polar cap radius (see, e.g., \citealt{LaPalombara12}).
The other solution (see Model I Smooth and Model III in Table~\ref{table:spectral_all}) returns a blackbody component with $kT\sim 0.1\,$keV and an emitting radius of the order of $10^3\,$km.
This is consistent with reprocessing by optically thick gas at the inner edge of the accretion disk, assumed to be at the magnetospheric radius $R_m\sim10^8\,$cm (see \citealt{Hickox04}).

The physical model \texttt{bwcyc} also fits the data adequately.
The best-fit configuration for the \texttt{bwcyc} spectral fit was obtained with the distance parameter frozen to the value of 14 kpc, thus favouring recent independent findings from \citet[]{Sharma22}.
The best-fit parameters are in agreement with those found for other similar sources \citep[]{Thalhammer21,Wolff16}.
Also, the relatively large value of the $\delta$ parameter confirms that the accretion process is dominated by bulk motion Comptonization.
Moreover, the best-fit accretion column radius, $r_0\sim90\,$m, is much smaller than the blackbody-emitting area ascribable to the NS surface (see above and Model I Deep and Model II in Table~\ref{table:spectral_all}). 
This result therefore disfavors the interpretation of the blackbody-emission component as originating from the polar cap.
Similarly to \citet[]{Thalhammer21}, we can use the similarity parameters $\xi$ and $\delta$ to derive the cross-sections $\Bar{\sigma}\sim3.5\times10^{-5}\sigma_T$ and $\sigma_\parallel\sim5\times10^{-6}\sigma_T$ (while the cross-section perpendicular to the magnetic field, $\sigma_\perp$, is fixed to the Thomson cross-section within the model).
The results are consistent with the expectation that $\sigma_\parallel<\Bar{\sigma}<\sigma_\perp$ \citep[]{Canuto71}.
Additionally, the model was able to fit the data with the magnetic field strength parameter linked to the value derived by the cyclotron line best-fit centroid energy, which supports the cyclotron line finding and the magnetic field value derived by it.

\subsection{A steady spectrum throughout the outburst}\label{subsec:discussion_spec2}

\begin{figure}[!t]
\includegraphics[width=.47\textwidth]{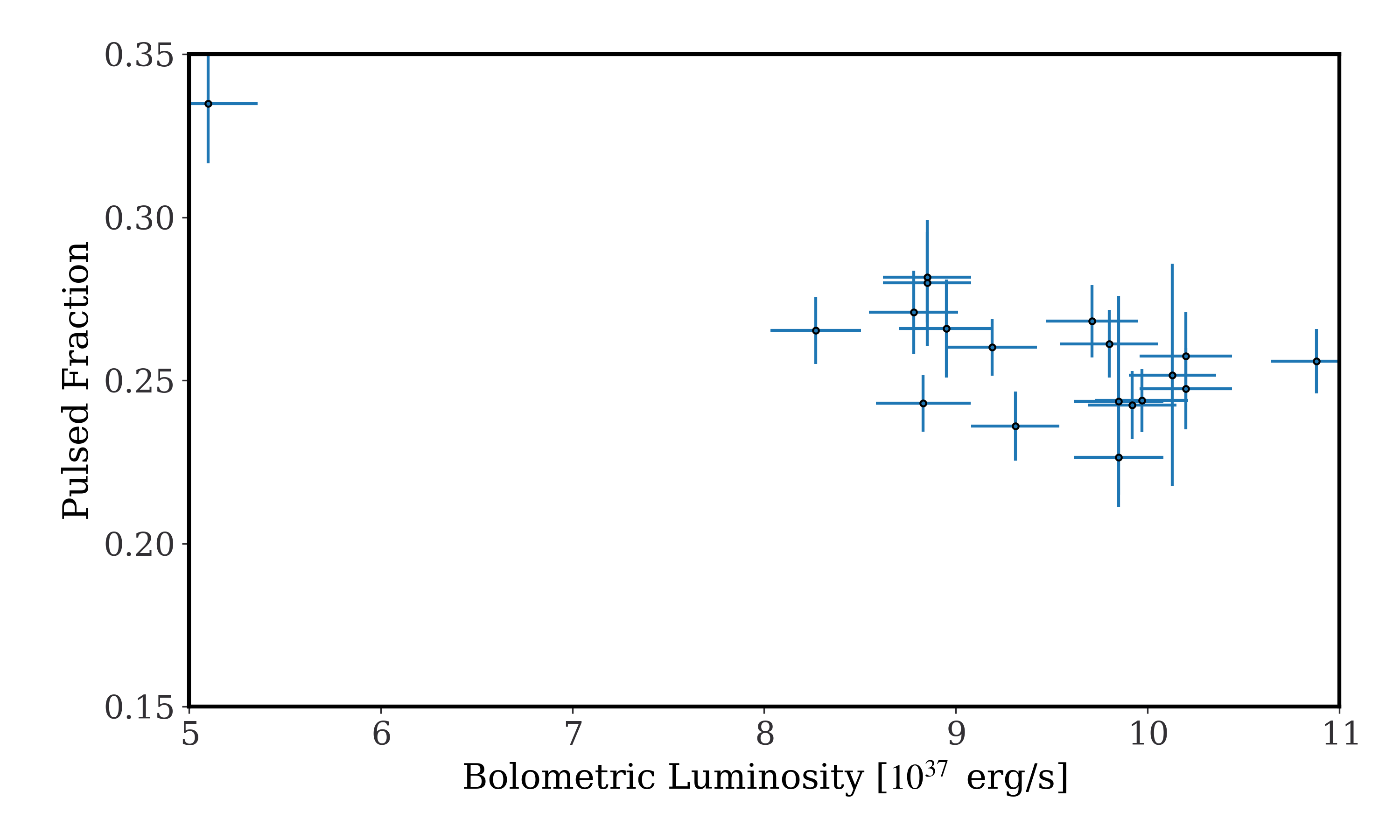}
\caption{Pulsed fraction (y-axis) as a function of luminosity (x-axis in units of $10^{37}\,$erg/s and for a distance of 14 kpc) for GRO J1750-27 as observed by \textit{NICER} along the outburst in 2021. \label{fig:pulsed_fraction_vs_luminosity}}
\end{figure}

The spectral analysis of single \textit{NICER} ObsIDs reveals a spectrum that is almost unchanging throughout the probed outburst stages (see Table~\ref{table:spectral_nicer}), both in its continuum (such as the photon index $\Gamma$), and in its discrete features (that is the Fe K$\alpha$ line best-fit parameters).

In XRPs, the iron line is nearly ubiquitously observed (see, e.g., \citealt[]{Nagase91}).
It is believed to be due to fluorescence emission resulting from reflection of the primary X-ray emission usually off the accretion disk (but also off the stellar wind or the surface of the donor star, \citealt[]{Basko1974, Inoue85}).
The iron line parameters observed in this work do not show any clear correlation with flux (see Table~\ref{table:spectral_nicer}).
Several other XRPs, on the contrary, show a clear dependence of the iron line flux with the observed flux from the X-ray source (see, e.g., \citealt[]{Reig+13, Jaisawal+19}).

The spectral steadiness is also confirmed by the model-independent investigation of the HR (see Fig.\,\ref{fig:total_plots}b).
This is contrary to other sources where, even for small luminosity variability (i.e., $\Delta L_X{\sim}10\%$), a hardening 
or softening of the spectrum is observed according to their accretion regime (see, e.g., \citealt[]{Klochkov+11, Reig+13, Fuerst+14, Malacaria+15}). 
Furthermore, distinct accretion regimes are separated by a certain critical luminosity \citep[]{Basko+Sunyaev76, Becker+12, Mushtukov15_crit_lum}, where the luminosity becomes strong enough that a radiation shock forms and stops the accretion flow above the NS surface.
As a reference, the critical luminosity obtained employing the model by \citet[see their Eq.~55]{Becker+12} is:
\begin{equation}\label{eq:BW}
\begin{split}    
    L^a_{\rm crit} =& 1.28\times10^{37}\,erg/s \left(\frac{\Lambda}{0.1}\right)^{-7/5} \left(\frac{M_{\rm NS}}{1.4M_\odot}\right)^{29/30}\\& \left(\frac{R_{\rm NS}}{10\,km}\right)^{1/10} \left(\frac{E_{\rm cyc}}{10\,\rm keV}\right)^{16/15},
\end{split}
\end{equation}

where $\Lambda$ is the accretion flow geometry constant, and M$_{\rm NS}$ and R$_{\rm NS}$ are the mass and radius of the NS, respectively.
Adopting $\Lambda= 0.1$ (for disk accretion), R$_{\rm NS}=12\,$km, M$_{\rm NS}{=}1.4\,\rm M_{\odot}$ and $E_{cyc}=44\,$keV in Eq.\,\ref{eq:BW} results in a critical luminosity of $L^a_{\rm crit}=6.3\times10^{37}\,$erg/s.
On the other hand, for a value of the cyclotron line energy of E$_{\rm cyc}=44\,$keV, \citet{Mushtukov15_crit_lum} predicts a critical luminosity value $L_{\rm crit}^b\sim2\times10^{37}\,$erg/s (see, e.g., their Fig.~7 for the case of pure X-mode polarization). 

However, \citet[]{Postnov+15} observed a flattening of the hardness ratio above the critical luminosity in several sources (i.e., 4U 0115+63, V 0332+53, EXO 2030+375, A 0535+26 and MXB 0656-072). \citet{Kuhnel17} also observed indications of a photon index break at high luminosity in GRO J1008-57 (see their Fig.~6), implying a possible constant spectrum above the critical luminosity.
The spectral results presented here in Table~\ref{table:spectral_nicer} and Fig.~\ref{fig:total_plots} are therefore consistent with the saturation effect of the spectral hardness described in \citet[]{Postnov+15}.
Their interpretation is based on the fraction of radiation reflected off the NS atmosphere with respect to the total emission.
Since Compton scattering is the most important process affecting the reflected radiation (see, e.g. \citealt[]{Poutanen13}), softer photons are absorbed while harder photons are scattered.
Therefore, the reflected radiation from the NS atmosphere is harder than its incident emission, and the spectrum becomes harder as the luminosity increases.
However, such a trend breaks down in the supercritical accretion regime, where the accretion column grows so high that the fraction of reflected radiation from the NS surface to the total emission becomes increasingly small, thus producing a flattening of the spectral HR.
As the GRO J1750-27 observations analyzed here were all carried out at a high-luminosity regime, and the observed HR remains remarkably flat, 
we might interpret the flat HR as the plateau observed by \citet[]{Postnov+15} and as due to the saturation effect at work in the supercritical regime.
This might imply that the accretion column has reached the maximum height imposed by the accretion physics limitations described in \citet{Poutanen13} and \citet{ Mushtukov15max}. Following their predictions, we infer a column height of $\sim2-5\,$km for the highest luminosity observed in this work from GRO~J1750-27.
Additional evidence supporting the saturation effect scenario is provided by the timing analysis and related discussion (Sect.~\ref{subsec:timing} and \ref{subsec:discussion_timing}, respectively).

\subsection{A steady pulse profile and pulsed fraction}\label{subsec:discussion_timing}

Our timing analysis of \src as observed by \textit{NICER} during the outburst in 2021 reveals a remarkably steady pulse profile (see Fig.s~\ref{fig:heatmap}, \ref{fig:profiles}) and pulsed fraction (see Fig.~\ref{fig:total_plots}c).
The pulse-profile heat map shows a dominant single peak at pulse phase ${\sim}1.0$.
Nonetheless, a transient feature appears at $\phi{\sim}0.6$ only for a few consecutive \textit{NICER} observations during the outburst rising stage, in the form of a secondary peak. The feature shows up not only at ObsIDs 09 and 10, where the secondary peak might be an artefact due to the short exposure times, but also at ObsID 11, where in fact it appears more prominently.
Moreover, a broadening of the main peak as the outburst evolves is also apparent in Figs.~\ref{fig:heatmap} and \ref{fig:profiles}.
According to our derived conversion factor $C_f$ (see Sect.~\ref{subsec:timing}), the bolometric luminosity corresponding to those ObsIDs is about $9(1)\times10^{37}\,$erg\,s$^{-1}$, while a peak luminosity of about $1.1(1)\times10^{38}\,$erg\,s$^{-1}$ is reached during ObsID 15.
However, the same feature is not present during the decline of the outburst, when a similar value of the luminosity is crossed again.
It is worth noting that a similar secondary peak, anticipating the main peak in phase by about $\sim0.3$, also appears in the \textit{NuSTAR} $9-18\,$keV pulse profile (see Fig. 5 from \citealt[]{Devaraj22}).
Apart from this transient feature, the pulse profiles do not show any appreciable luminosity dependence.
This is contrary to several other sources (see, e.g., \citealt[]{Klochkov08,Malacaria+15, Epili+17, Koliopanos+18, Wilson-Hodge18, Wang22}), where a drastic change of the pulse profile is observed as a function of luminosity, and is interpreted as a change of the accretion column beaming pattern \citep[]{Basko+Sunyaev76, Becker+12, Mushtukov15_crit_lum}.
Here, despite the probed outburst stages ranging from a bolometric luminosity of about $5\times10^{37}\,$erg\,s$^{-1}$ to a peak luminosity that is about $60\%$ of the Eddington luminosity for a 1.4\,M$_\odot$ NS (see Fig.~\ref{fig:heatmap}), the pulse profile remains almost unchanged, suggesting that the source is most likely always observed by NICER in the same, supercritical accretion regime.
This also can be interpreted as an indication of the saturation effect at work in the supercritical regime as indicated by \citet[]{Postnov+15} and already invoked in Sect.\,\ref{subsec:discussion_spec2}.
In fact, at the probed high-luminosity values the pulse profile is not affected by dramatic beaming pattern changes or by emission reflected off the NS surface, which might alter its overall shape. As a consequence, the pulse profile appears almost unaltered throughout the observed outburst stages.

Finally, we notice that the PF behavior also  seems to follow a similar saturation effect to that shown by the spectral HR (see Sect.~\ref{subsec:spectral}).
Indeed, the PF shows only a minor decrease during the rising part of the outburst, and then flattens around $25\%$ (see Fig.~\ref{fig:total_plots}c).
This suggests a similar interpretation where, above a certain flux level, the beaming pattern contributions do not significantly change as a function of the accretion rate, nor does the radiation reflected off the NS surface affect the PF at any appreciable level. 
Moreover, we report the pulsed-fraction variability as a function of luminosity in Fig.~\ref{fig:pulsed_fraction_vs_luminosity}. Several XRPs have been observed to show a luminosity-dependent PF, which is often interpreted in terms of beaming pattern and geometrical changes \citep[]{Lutovinov09,Yang2018, Gupta2019, Lutovinov21}. The relatively large error bars and lack of data for intermediate-luminosity stages in Fig.~\ref{fig:pulsed_fraction_vs_luminosity} prevent us from drawing any firm conclusions. However, except for the lowest luminosity data point, the data suggest a PF that is insensitive to the observed bolometric luminosity.

\section{Summary}

We performed spectral and timing analysis of the accreting XRP GRO J1750-27 as observed by \textit{NICER} and \textit{NuSTAR} at the peak of its latest outburst toward the end of 2021.
Our main results can be summarized as follows:

\begin{itemize}
    \item We tested unexplored spectral models and confirm the discovery of a cyclotron line from this source at about 44 keV, implying a magnetic field strength of ${\sim}4.7\times10^{12}\,$G (consistent with recent independent findings). The cyclotron line depth is possibly amongst the largest ever observed. Our results also highlight a highly absorbed source ($N_H\sim(5-8)\times10^{22}\,$cm\,$^{-2}$), and a soft blackbody component likely originating from reprocessing at the inner edge of the accretion disk.
    \item The spectral data can be fitted with a physical model based on bulk and thermal Comptonization (\texttt{bwcyc}), favoring a distance of 14 kpc and supporting the magnetic field strength derived by the cyclotron line.
    This model also features a bulk-Comptonization-dominated spectrum, a relatively narrow accretion column, and a scattering cross-section parallel to the magnetic field that is severely reduced relative to the Thomson cross-section.
    \item The \textit{NICER} monitoring shows an almost immutable spectrum throughout all observations. This is interpreted here as a saturation effect occurring above the critical luminosity and due to emission from the accretion column reflected off the surface of the NS \citep[]{Postnov+15}.
    \item The 0.5-10 keV pulse profiles as monitored by \textit{NICER} show a simple one-peaked shape that remains almost identical throughout the outburst. The hint of a secondary peak appears at an intermediate luminosity during the outburst rise (but not during the decay). This is similar to what is observed by \textit{NuSTAR} at harder energies (i.e., 9-18 keV).
    \item The pulsed fraction also appears to be independent of time and luminosity, supporting the saturation effect interpretation.
\end{itemize}
 
To the best of our knowledge, this might be the first time that the saturation effect is observed in the spectral properties, the pulsed fraction, and the pulse profile shape at the same time.
Given the elusive nature of several physical effects at work in GRO J1750-27, of which the present analysis reveals only a mere glimpse, an investigation into the luminosity dependence of the cyclotron line at lower accretion rates is encouraged, together with an X-ray high- and low-luminosity spectropolarimetry study, which could help us to discern the accretion column beaming pattern and the amount of radiation reflected off the surface of the NS.

\begin{acknowledgements}
\balance

This research has made use of data and software provided by the High Energy Astrophysics Science Archive Research Center (HEASARC), which is a service of the Astrophysics Science Division at NASA/GSFC and the High Energy Astrophysics Division of the Smithsonian Astrophysical Observatory. We acknowledge extensive use of the NASA Abstract Database Service (ADS).
This work used data from the NuSTAR mission, a project led by the California Institute of Technology, managed by the JPL, and funded by NASA, and has utilised the NUSTARDAS software package, jointly developed by the ASI Science Data Center, Italy, and the California Institute of Technology, USA.
DA acknowledges support from the Royal Society.
This work was supported by NASA through the NICER mission and the Astrophysics Explorers Program.
\end{acknowledgements}

\bibliographystyle{yahapj}
\bibliography{references}

\begin{thebibliography}{}
\providecommand\natexlab[1]{#1}
\providecommand\JournalTitle[1]{#1}

\bibitem[{{Aftab} {et~al.}(2019){Aftab}, {Paul}, \& {Kretschmar}}]{Aftab19}
{Aftab}, N., {Paul}, B., \& {Kretschmar}, P. 2019,
  \href{http://dx.doi.org/10.3847/1538-4365/ab2a77}{\JournalTitle{\apjs}, 243,
  29}

\bibitem[{{Arnaud}(1996)}]{Arnaud96}
{Arnaud}, K.~A. 1996, in Astronomical Society of the Pacific Conference Series,
  Vol. 101, Astronomical Data Analysis Software and Systems V, ed. G.~H.
  {Jacoby} \& J.~{Barnes}, 17

\bibitem[{{Bailer-Jones} {et~al.}(2021){Bailer-Jones}, {Rybizki}, {Fouesneau},
  {Demleitner}, \& {Andrae}}]{Bailer-Jones21}
{Bailer-Jones}, C.~A.~L., {Rybizki}, J., {Fouesneau}, M., {Demleitner}, M., \&
  {Andrae}, R. 2021,
  \href{http://dx.doi.org/10.3847/1538-3881/abd806}{\JournalTitle{\aj}, 161,
  147}

\bibitem[{{Basko} \& {Sunyaev}(1976)}]{Basko+Sunyaev76}
{Basko}, M.~M., \& {Sunyaev}, R.~A. 1976,
  \href{http://dx.doi.org/10.1093/mnras/175.2.395}{\JournalTitle{\mnras}, 175,
  395}

\bibitem[{{Basko} {et~al.}(1974){Basko}, {Sunyaev}, \& {Titarchuk}}]{Basko1974}
{Basko}, M.~M., {Sunyaev}, R.~A., \& {Titarchuk}, L.~G. 1974,
  \JournalTitle{\aap}, 31, 249

\bibitem[{{Becker} \& {Wolff}(2007)}]{Becker+Wolff07}
{Becker}, P.~A., \& {Wolff}, M.~T. 2007,
  \href{http://dx.doi.org/10.1086/509108}{\JournalTitle{\apj}, 654, 435}

\bibitem[{{Becker} {et~al.}(2012){Becker}, {Klochkov}, {Sch{\"o}nherr},
  {Nishimura}, {Ferrigno}, {Caballero}, {Kretschmar}, {Wolff}, {Wilms}, \&
  {Staubert}}]{Becker+12}
{Becker}, P.~A., {Klochkov}, D., {Sch{\"o}nherr}, G., {et~al.} 2012,
  \href{http://dx.doi.org/10.1051/0004-6361/201219065}{\JournalTitle{\aap},
  544, A123}

\bibitem[{{Boissay} {et~al.}(2015){Boissay}, {Chenevez}, {Wilms}, {Grinberg},
  {Del Santo}, {Bazzano}, {Capitanio}, {Tarana}, {Paizis}, {Watanabe},
  {Rodriguez}, {Goetz}, {Kuulkers}, \& {Ferrigno}}]{Boissay15}
{Boissay}, R., {Chenevez}, J., {Wilms}, J., {et~al.} 2015, \JournalTitle{The
  Astronomer's Telegram}, 7096, 1

\bibitem[{{Boldin} {et~al.}(2013){Boldin}, {Tsygankov}, \&
  {Lutovinov}}]{Boldin13}
{Boldin}, P.~A., {Tsygankov}, S.~S., \& {Lutovinov}, A.~A. 2013,
  \href{http://dx.doi.org/10.1134/S1063773713060029}{\JournalTitle{Astronomy
  Letters}, 39, 375}

\bibitem[{{Canuto} {et~al.}(1971){Canuto}, {Lodenquai}, \&
  {Ruderman}}]{Canuto71}
{Canuto}, V., {Lodenquai}, J., \& {Ruderman}, M. 1971,
  \href{http://dx.doi.org/10.1103/PhysRevD.3.2303}{\JournalTitle{\prd}, 3,
  2303}

\bibitem[{{Cash}(1979)}]{Cash79}
{Cash}, W. 1979, \href{http://dx.doi.org/10.1086/156922}{\JournalTitle{\apj},
  228, 939}

\bibitem[{{Coburn} {et~al.}(2002){Coburn}, {Heindl}, {Rothschild}, {Gruber},
  {Kreykenbohm}, {Wilms}, {Kretschmar}, \& {Staubert}}]{Coburn+02}
{Coburn}, W., {Heindl}, W.~A., {Rothschild}, R.~E., {et~al.} 2002,
  \href{http://dx.doi.org/10.1086/343033}{\JournalTitle{\apj}, 580, 394}

\bibitem[{{Devaraj} \& {Paul}(2022)}]{Devaraj22}
{Devaraj}, A., \& {Paul}, B. 2022,
  \href{http://dx.doi.org/10.1093/mnrasl/slac052}{\JournalTitle{\mnras}, 514,
  L46}

\bibitem[{{Epili} {et~al.}(2017){Epili}, {Naik}, {Jaisawal}, \&
  {Gupta}}]{Epili+17}
{Epili}, P., {Naik}, S., {Jaisawal}, G.~K., \& {Gupta}, S. 2017,
  \href{http://dx.doi.org/10.1093/mnras/stx2247}{\JournalTitle{\mnras}, 472,
  3455}

\bibitem[{{Fabricius} {et~al.}(2021){Fabricius}, {Luri}, {Arenou}, {Babusiaux},
  {Helmi}, {Muraveva}, {Reyl{\'e}}, {Spoto}, {Vallenari}, {Antoja}, {Balbinot},
  {Barache}, {Bauchet}, {Bragaglia}, {Busonero}, {Cantat-Gaudin}, {Carrasco},
  {Diakit{\'e}}, {Fabrizio}, {Figueras}, {Garcia-Gutierrez}, {Garofalo},
  {Jordi}, {Kervella}, {Khanna}, {Leclerc}, {Licata}, {Lambert}, {Marrese},
  {Masip}, {Ramos}, {Robichon}, {Robin}, {Romero-G{\'o}mez}, {Rubele}, \&
  {Weiler}}]{Fabricius21}
{Fabricius}, C., {Luri}, X., {Arenou}, F., {et~al.} 2021,
  \href{http://dx.doi.org/10.1051/0004-6361/202039834}{\JournalTitle{\aap},
  649, A5}

\bibitem[{{Farinelli} {et~al.}(2012){Farinelli}, {Ceccobello}, {Romano}, \&
  {Titarchuk}}]{Farinelli12}
{Farinelli}, R., {Ceccobello}, C., {Romano}, P., \& {Titarchuk}, L. 2012,
  \href{http://dx.doi.org/10.1051/0004-6361/201118008}{\JournalTitle{\aap},
  538, A67}

\bibitem[{{Ferrigno} {et~al.}(2009){Ferrigno}, {Becker}, {Segreto}, {Mineo}, \&
  {Santangelo}}]{Ferrigno09}
{Ferrigno}, C., {Becker}, P.~A., {Segreto}, A., {Mineo}, T., \& {Santangelo},
  A. 2009,
  \href{http://dx.doi.org/10.1051/0004-6361/200809373}{\JournalTitle{\aap},
  498, 825}

\bibitem[{{Ferrigno} {et~al.}(2019){Ferrigno}, {Bozzo}, {Sanna}, {Jaisawal},
  {Girard}, {Di Salvo}, \& {Burderi}}]{Ferrigno2019}
{Ferrigno}, C., {Bozzo}, E., {Sanna}, A., {et~al.} 2019,
  \href{http://dx.doi.org/10.1051/0004-6361/201935185}{\JournalTitle{\aap},
  624, A142}

\bibitem[{{Ferrigno} {et~al.}(2007){Ferrigno}, {Segreto}, {Santangelo},
  {Wilms}, {Kreykenbohm}, {Denis}, \& {Staubert}}]{Ferrigno07}
{Ferrigno}, C., {Segreto}, A., {Santangelo}, A., {et~al.} 2007,
  \href{http://dx.doi.org/10.1051/0004-6361:20053878}{\JournalTitle{\aap}, 462,
  995}

\bibitem[{{Finger} \& {Wilson-Hodge}(2014)}]{Finger+Wilson14}
{Finger}, M.~H., \& {Wilson-Hodge}, C.~A. 2014, \JournalTitle{ATel}, 6839, 0

\bibitem[{{F{\"u}rst} {et~al.}(2014){F{\"u}rst}, {Pottschmidt}, {Wilms},
  {Tomsick}, {Bachetti}, {Boggs}, {Christensen}, {Craig}, {Grefenstette},
  {Hailey}, {Harrison}, {Madsen}, {Miller}, {Stern}, {Walton}, \&
  {Zhang}}]{Fuerst+14}
{F{\"u}rst}, F., {Pottschmidt}, K., {Wilms}, J., {et~al.} 2014,
  \href{http://dx.doi.org/10.1088/0004-637X/780/2/133}{\JournalTitle{\apj},
  780, 133}

\bibitem[{{Gendreau} \& {Arzoumanian}(2017)}]{Gendreau2017}
{Gendreau}, K., \& {Arzoumanian}, Z. 2017,
  \href{http://dx.doi.org/10.1038/s41550-017-0301-3}{\JournalTitle{Nature
  Astronomy}, 1, 895}

\bibitem[{{Grove} {et~al.}(1995){Grove}, {Strickman}, {Johnson}, {Kurfess},
  {Kinzer}, {Starr}, {Jung}, {Kendziorra}, {Kretschmar}, {Maisack}, \&
  {Staubert}}]{Grove95}
{Grove}, J.~E., {Strickman}, M.~S., {Johnson}, W.~N., {et~al.} 1995,
  \href{http://dx.doi.org/10.1086/187706}{\JournalTitle{\apjl}, 438, L25}

\bibitem[{{Gupta} {et~al.}(2019){Gupta}, {Naik}, \& {Jaisawal}}]{Gupta2019}
{Gupta}, S., {Naik}, S., \& {Jaisawal}, G.~K. 2019,
  \href{http://dx.doi.org/10.1093/mnras/stz2795}{\JournalTitle{\mnras}, 490,
  2458}

\bibitem[{{Harrison} {et~al.}(2013){Harrison}, {Craig}, {Christensen},
  {Hailey}, {Zhang}, {Boggs}, {Stern}, {Cook}, {Forster}, {Giommi},
  {Grefenstette}, {Kim}, {Kitaguchi}, {Koglin}, {Madsen}, {Mao}, {Miyasaka},
  {Mori}, {Perri}, {Pivovaroff}, {Puccetti}, {Rana}, {Westergaard}, {Willis},
  {Zoglauer}, {An}, {Bachetti}, {Barri{\`e}re}, {Bellm}, {Bhalerao},
  {Brejnholt}, {Fuerst}, {Liebe}, {Markwardt}, {Nynka}, {Vogel}, {Walton},
  {Wik}, {Alexander}, {Cominsky}, {Hornschemeier}, {Hornstrup}, {Kaspi},
  {Madejski}, {Matt}, {Molendi}, {Smith}, {Tomsick}, {Ajello}, {Ballantyne},
  {Balokovi{\'c}}, {Barret}, {Bauer}, {Blandford}, {Brandt}, {Brenneman},
  {Chiang}, {Chakrabarty}, {Chenevez}, {Comastri}, {Dufour}, {Elvis}, {Fabian},
  {Farrah}, {Fryer}, {Gotthelf}, {Grindlay}, {Helfand}, {Krivonos}, {Meier},
  {Miller}, {Natalucci}, {Ogle}, {Ofek}, {Ptak}, {Reynolds}, {Rigby},
  {Tagliaferri}, {Thorsett}, {Treister}, \& {Urry}}]{Harrison13}
{Harrison}, F.~A., {Craig}, W.~W., {Christensen}, F.~E., {et~al.} 2013,
  \href{http://dx.doi.org/10.1088/0004-637X/770/2/103}{\JournalTitle{\apj},
  770, 103}

\bibitem[{{HI4PI Collaboration} {et~al.}(2016){HI4PI Collaboration}, {Ben
  Bekhti}, {Fl{\"o}er}, {Keller}, {Kerp}, {Lenz}, {Winkel}, {Bailin},
  {Calabretta}, {Dedes}, {Ford}, {Gibson}, {Haud}, {Janowiecki}, {Kalberla},
  {Lockman}, {McClure-Griffiths}, {Murphy}, {Nakanishi}, {Pisano}, \&
  {Staveley-Smith}}]{HI4PI2016}
{HI4PI Collaboration}, {Ben Bekhti}, N., {Fl{\"o}er}, L., {et~al.} 2016,
  \href{http://dx.doi.org/10.1051/0004-6361/201629178}{\JournalTitle{\aap},
  594, A116}

\bibitem[{{Hickox} {et~al.}(2004){Hickox}, {Narayan}, \& {Kallman}}]{Hickox04}
{Hickox}, R.~C., {Narayan}, R., \& {Kallman}, T.~R. 2004,
  \href{http://dx.doi.org/10.1086/423928}{\JournalTitle{\apj}, 614, 881}

\bibitem[{{Inoue}(1985)}]{Inoue85}
{Inoue}, H. 1985,
  \href{http://dx.doi.org/10.1007/BF00212905}{\JournalTitle{\ssr}, 40, 317}

\bibitem[{Iyer {et~al.}(2015)Iyer, Mukherjee, Dewangan, Bhattacharya, \&
  Seetha}]{Iyer15}
Iyer, N., Mukherjee, D., Dewangan, G.~C., Bhattacharya, D., \& Seetha, S. 2015,
  \href{http://dx.doi.org/10.1093/mnras/stv1942}{\JournalTitle{MNRAS}, 454,
  741}

\bibitem[{{Jaisawal} \& {Naik}(2016)}]{Jaisawal2016}
{Jaisawal}, G.~K., \& {Naik}, S. 2016,
  \href{http://dx.doi.org/10.1093/mnrasl/slw108}{\JournalTitle{\mnras}, 461,
  L97}

\bibitem[{{Jaisawal} {et~al.}(2019){Jaisawal}, {Wilson-Hodge}, {Fabian},
  {Naik}, {Chakrabarty}, {Kretschmar}, {Ballantyne}, {Ludlam}, {Chenevez},
  {Altamirano}, {Arzoumanian}, {F{\"u}rst}, {Gendreau}, {Guillot}, {Malacaria},
  {Miller}, {Stevens}, \& {Wolff}}]{Jaisawal+19}
{Jaisawal}, G.~K., {Wilson-Hodge}, C.~A., {Fabian}, A.~C., {et~al.} 2019,
  \href{http://dx.doi.org/10.3847/1538-4357/ab4595}{\JournalTitle{\apj}, 885,
  18}

\bibitem[{{Klochkov} {et~al.}(2008){Klochkov}, {Santangelo}, {Staubert}, \&
  {Ferrigno}}]{Klochkov08}
{Klochkov}, D., {Santangelo}, A., {Staubert}, R., \& {Ferrigno}, C. 2008,
  \href{http://dx.doi.org/10.1051/0004-6361:200810673}{\JournalTitle{\aap},
  491, 833}

\bibitem[{{Klochkov} {et~al.}(2011){Klochkov}, {Staubert}, {Santangelo},
  {Rothschild}, \& {Ferrigno}}]{Klochkov+11}
{Klochkov}, D., {Staubert}, R., {Santangelo}, A., {Rothschild}, R.~E., \&
  {Ferrigno}, C. 2011,
  \href{http://dx.doi.org/10.1051/0004-6361/201116800}{\JournalTitle{\aap},
  532, A126}

\bibitem[{{Klochkov} {et~al.}(2007){Klochkov}, {Horns}, {Santangelo},
  {Staubert}, {Segreto}, {Ferrigno}, {Kretschmar}, {Kreykenbohm}, {La Barbera},
  {Masetti}, {McCollough}, {Pottschmidt}, {Sch{\"o}nherr}, \&
  {Wilms}}]{Klochkov07}
{Klochkov}, D., {Horns}, D., {Santangelo}, A., {et~al.} 2007,
  \href{http://dx.doi.org/10.1051/0004-6361:20066801}{\JournalTitle{\aap}, 464,
  L45}

\bibitem[{{Koliopanos} \& {Vasilopoulos}(2018)}]{Koliopanos+18}
{Koliopanos}, F., \& {Vasilopoulos}, G. 2018, \JournalTitle{ArXiv e-prints},
  \href{http://arxiv.org/abs/1801.01168}{{\sffamily arXiv:1801.01168
  [astro-ph.HE]}}

\bibitem[{{Kong} {et~al.}(2022){Kong}, {Zhang}, {Zhang}, {Ji}, {Doroshenko},
  {Santangelo}, {Chen}, {Lu}, {Ge}, {Wang}, {Tao}, {Qu}, {Li}, {Liu}, {Liao},
  {Chang}, {Peng}, \& {Shui}}]{Kong2022}
{Kong}, L.-D., {Zhang}, S., {Zhang}, S.-N., {et~al.} 2022,
  \href{http://dx.doi.org/10.3847/2041-8213/ac7711}{\JournalTitle{\apjl}, 933,
  L3}

\bibitem[{{Krimm} {et~al.}(2008){Krimm}, {Barthelmy}, {Cummings}, {Fenimore},
  {Gehrels}, {Markwardt}, {Palmer}, {Parsons}, {Sakamoto}, {Sato}, {Skinner},
  {Stamatikos}, \& {Tueller}}]{Krimm08}
{Krimm}, H.~A., {Barthelmy}, S.~D., {Cummings}, J., {et~al.} 2008,
  \JournalTitle{The Astronomer's Telegram}, 1376, 1

\bibitem[{{K{\"u}hnel} {et~al.}(2017){K{\"u}hnel}, {F{\"u}rst}, {Pottschmidt},
  {Kreykenbohm}, {Ballhausen}, {Falkner}, {Rothschild}, {Klochkov}, \&
  {Wilms}}]{Kuhnel17}
{K{\"u}hnel}, M., {F{\"u}rst}, F., {Pottschmidt}, K., {et~al.} 2017,
  \href{http://dx.doi.org/10.1051/0004-6361/201629131}{\JournalTitle{\aap},
  607, A88}

\bibitem[{{La Palombara} {et~al.}(2012){La Palombara}, {Sidoli}, {Esposito},
  {Tiengo}, \& {Mereghetti}}]{LaPalombara12}
{La Palombara}, N., {Sidoli}, L., {Esposito}, P., {Tiengo}, A., \&
  {Mereghetti}, S. 2012,
  \href{http://dx.doi.org/10.1051/0004-6361/201118221}{\JournalTitle{\aap},
  539, A82}

\bibitem[{{LaMarr} {et~al.}(2016){LaMarr}, {Prigozhin}, {Remillard}, {Malonis},
  {Gendreau}, {Arzoumanian}, {Markwardt}, \& {Baumgartner}}]{LaMarr2016}
{LaMarr}, B., {Prigozhin}, G., {Remillard}, R., {et~al.} 2016,
  \href{http://dx.doi.org/10.1117/12.2232784}{in Society of Photo-Optical
  Instrumentation Engineers (SPIE) Conference Series, Vol. 9905, Space
  Telescopes and Instrumentation 2016: Ultraviolet to Gamma Ray, ed. J.-W.~A.
  {den Herder}, T.~{Takahashi}, \& M.~{Bautz}}, 99054W

\bibitem[{{Leahy} {et~al.}(1983){Leahy}, {Darbro}, {Elsner}, {Weisskopf},
  {Kahn}, {Sutherland}, \& {Grindlay}}]{Leahy+83}
{Leahy}, D.~A., {Darbro}, W., {Elsner}, R.~F., {et~al.} 1983,
  \href{http://dx.doi.org/10.1086/160766}{\JournalTitle{\apj}, 266, 160}

\bibitem[{{Lutovinov} {et~al.}(2012){Lutovinov}, {Tsygankov}, \&
  {Chernyakova}}]{Lutovinov12}
{Lutovinov}, A., {Tsygankov}, S., \& {Chernyakova}, M. 2012,
  \href{http://dx.doi.org/10.1111/j.1365-2966.2012.21036.x}{\JournalTitle{\mnras},
  423, 1978}

\bibitem[{{Lutovinov} {et~al.}(2021){Lutovinov}, {Tsygankov}, {Molkov},
  {Doroshenko}, {Mushtukov}, {Arefiev}, {Lapshov}, {Tkachenko}, \&
  {Pavlinsky}}]{Lutovinov21}
{Lutovinov}, A., {Tsygankov}, S., {Molkov}, S., {et~al.} 2021,
  \href{http://dx.doi.org/10.3847/1538-4357/abec43}{\JournalTitle{\apj}, 912,
  17}

\bibitem[{{Lutovinov} \& {Tsygankov}(2009)}]{Lutovinov09}
{Lutovinov}, A.~A., \& {Tsygankov}, S.~S. 2009,
  \href{http://dx.doi.org/10.1134/S1063773709070019}{\JournalTitle{Astronomy
  Letters}, 35, 433}

\bibitem[{{Lutovinov} {et~al.}(2019){Lutovinov}, {Tsygankov}, {Karasev},
  {Molkov}, \& {Doroshenko}}]{Lutovinov19}
{Lutovinov}, A.~A., {Tsygankov}, S.~S., {Karasev}, D.~I., {Molkov}, S.~V., \&
  {Doroshenko}, V. 2019,
  \href{http://dx.doi.org/10.1093/mnras/stz437}{\JournalTitle{\mnras}, 485,
  770}

\bibitem[{{Madsen} {et~al.}(2021){Madsen}, {Forster}, {Grefenstette},
  {Harrison}, \& {Miyasaka}}]{Madsen21}
{Madsen}, K.~K., {Forster}, K., {Grefenstette}, B.~W., {Harrison}, F.~A., \&
  {Miyasaka}, H. 2021, \JournalTitle{arXiv e-prints}, arXiv:2110.11522

\bibitem[{{Madsen} {et~al.}(2015){Madsen}, {Harrison}, {Markwardt}, {An},
  {Grefenstette}, {Bachetti}, {Miyasaka}, {Kitaguchi}, {Bhalerao}, {Boggs},
  {Christensen}, {Craig}, {Forster}, {Fuerst}, {Hailey}, {Perri}, {Puccetti},
  {Rana}, {Stern}, {Walton}, {J{\o}rgen Westergaard}, \& {Zhang}}]{Madsen15}
{Madsen}, K.~K., {Harrison}, F.~A., {Markwardt}, C.~B., {et~al.} 2015,
  \href{http://dx.doi.org/10.1088/0067-0049/220/1/8}{\JournalTitle{\apjs}, 220,
  8}

\bibitem[{{Malacaria} {et~al.}(2020){Malacaria}, {Jenke}, {Roberts},
  {Wilson-Hodge}, {Cleveland}, {Mailyan}, \& {GBM Accreting Pulsars Program
  Team}}]{Malacaria20}
{Malacaria}, C., {Jenke}, P., {Roberts}, O.~J., {et~al.} 2020,
  \href{http://dx.doi.org/10.3847/1538-4357/ab855c}{\JournalTitle{\apj}, 896,
  90}

\bibitem[{{Malacaria} {et~al.}(2021{\natexlab{a}}){Malacaria}, {Jenke}, \&
  {Wilson-Hodge}}]{Malacaria21_AtelGRO}
{Malacaria}, C., {Jenke}, P., \& {Wilson-Hodge}, C. 2021{\natexlab{a}},
  \JournalTitle{The Astronomer's Telegram}, 14930, 1

\bibitem[{{Malacaria} {et~al.}(2015){Malacaria}, {Klochkov}, {Santangelo}, \&
  {Staubert}}]{Malacaria+15}
{Malacaria}, C., {Klochkov}, D., {Santangelo}, A., \& {Staubert}, R. 2015,
  \href{http://dx.doi.org/10.1051/0004-6361/201526417}{\JournalTitle{\aap},
  581, A121}

\bibitem[{{Malacaria} {et~al.}(2021{\natexlab{b}}){Malacaria}, {Kretschmar},
  {Madsen}, {Wilson-Hodge}, {Coley}, {Jenke}, {Lutovinov}, {Pottschmidt},
  {Tsygankov}, \& {Wilms}}]{Malacaria21}
{Malacaria}, C., {Kretschmar}, P., {Madsen}, K.~K., {et~al.}
  2021{\natexlab{b}},
  \href{http://dx.doi.org/10.3847/1538-4357/abddbc}{\JournalTitle{\apj}, 909,
  153}

\bibitem[{{Malacaria} {et~al.}(2022{\natexlab{a}}){Malacaria}, {Bhargava},
  {Coley}, {Ducci}, {Pradhan}, {Ballhausen}, {Fuerst}, {Islam}, {Jaisawal},
  {Jenke}, {Kretschmar}, {Kreykenbohm}, {Pottschmidt}, {Sokolova-Lapa},
  {Staubert}, {Wilms}, {Wilson-Hodge}, \& {Wolff}}]{Malacaria2022}
{Malacaria}, C., {Bhargava}, Y., {Coley}, J.~B., {et~al.} 2022{\natexlab{a}},
  \href{http://dx.doi.org/10.3847/1538-4357/ac524f}{\JournalTitle{\apj}, 927,
  194}

\bibitem[{{Malacaria} {et~al.}(2022{\natexlab{b}}){Malacaria}, {Coley},
  {Ducci}, {Fuerst}, {Jaisawal}, {Kretschmar}, {Pottschmidt}, {Pradhan}, \&
  {Wilms}}]{MalacariaAtel2022}
{Malacaria}, C., {Coley}, J.~B., {Ducci}, L., {et~al.} 2022{\natexlab{b}},
  \JournalTitle{The Astronomer's Telegram}, 15241, 1

\bibitem[{{Mihara} {et~al.}(1998){Mihara}, {Makishima}, \& {Nagase}}]{Mihara98}
{Mihara}, T., {Makishima}, K., \& {Nagase}, F. 1998,
  \href{http://dx.doi.org/10.1016/S0273-1177(98)00128-8}{\JournalTitle{Advances
  in Space Research}, 22, 987}

\bibitem[{{Mowlavi} {et~al.}(2006){Mowlavi}, {Kreykenbohm}, {Shaw},
  {Pottschmidt}, {Wilms}, {Rodriguez}, {Produit}, {Soldi}, {Larsson}, \&
  {Dubath}}]{Mowlavi06}
{Mowlavi}, N., {Kreykenbohm}, I., {Shaw}, S.~E., {et~al.} 2006,
  \href{http://dx.doi.org/10.1051/0004-6361:20054235}{\JournalTitle{\aap}, 451,
  187}

\bibitem[{{Mushtukov} \& {Tsygankov}(2022)}]{Mushtukov22}
{Mushtukov}, A., \& {Tsygankov}, S. 2022, \JournalTitle{arXiv e-prints},
  arXiv:2204.14185

\bibitem[{{Mushtukov} {et~al.}(2015{\natexlab{a}}){Mushtukov}, {Suleimanov},
  {Tsygankov}, \& {Poutanen}}]{Mushtukov15max}
{Mushtukov}, A.~A., {Suleimanov}, V.~F., {Tsygankov}, S.~S., \& {Poutanen}, J.
  2015{\natexlab{a}},
  \href{http://dx.doi.org/10.1093/mnras/stv2087}{\JournalTitle{\mnras}, 454,
  2539}

\bibitem[{{Mushtukov} {et~al.}(2015{\natexlab{b}}){Mushtukov}, {Suleimanov},
  {Tsygankov}, \& {Poutanen}}]{Mushtukov15_crit_lum}
---. 2015{\natexlab{b}},
  \href{http://dx.doi.org/10.1093/mnras/stu2484}{\JournalTitle{\mnras}, 447,
  1847}

\bibitem[{{Nagase}(1991)}]{Nagase91}
{Nagase}, F. 1991, \href{http://dx.doi.org/10.1007/BFb0031280}{in Iron Line
  Diagnostics in X-ray Sources, ed. A.~{Treves}, G.~C. {Perola}, \&
  L.~{Stella}, Vol. 385}, 111

\bibitem[{Okajima {et~al.}(2016)Okajima, Soong, Balsamo, Enoto, Olsen,
  Koenecke, Lozipone, Kearney, Fitzsimmons, Numata, Kenyon, Arzoumanian, \&
  Gendreau}]{Okajima2016}
Okajima, T., Soong, Y., Balsamo, E.~R., {et~al.} 2016,
  \href{http://dx.doi.org/10.1117/12.2234436}{in Space Telescopes and
  Instrumentation 2016: Ultraviolet to Gamma Ray, ed. J.-W.~A. den Herder,
  T.~Takahashi, \& M.~Bautz, Vol. 9905}, International Society for Optics and
  Photonics (SPIE), 1495

\bibitem[{{Postnov} {et~al.}(2015){Postnov}, {Gornostaev}, {Klochkov},
  {Laplace}, {Lukin}, \& {Shakura}}]{Postnov+15}
{Postnov}, K.~A., {Gornostaev}, M.~I., {Klochkov}, D., {et~al.} 2015,
  \href{http://dx.doi.org/10.1093/mnras/stv1393}{\JournalTitle{\mnras}, 452,
  1601}

\bibitem[{{Pottschmidt} {et~al.}(2005){Pottschmidt}, {Kreykenbohm}, {Wilms},
  {Coburn}, {Rothschild}, {Kretschmar}, {McBride}, {Suchy}, \&
  {Staubert}}]{Pottschmidt05}
{Pottschmidt}, K., {Kreykenbohm}, I., {Wilms}, J., {et~al.} 2005,
  \href{http://dx.doi.org/10.1086/498689}{\JournalTitle{\apjl}, 634, L97}

\bibitem[{{Poutanen} {et~al.}(2013){Poutanen}, {Mushtukov}, {Suleimanov},
  {Tsygankov}, {Nagirner}, {Doroshenko}, \& {Lutovinov}}]{Poutanen13}
{Poutanen}, J., {Mushtukov}, A.~A., {Suleimanov}, V.~F., {et~al.} 2013,
  \href{http://dx.doi.org/10.1088/0004-637X/777/2/115}{\JournalTitle{\apj},
  777, 115}

\bibitem[{Prigozhin {et~al.}(2016)Prigozhin, Gendreau, Doty, Foster, Remillard,
  Malonis, LaMarr, Vezie, Egan, Villasenor, Arzoumanian, Baumgartner, Scholze,
  Laubis, Krumrey, \& Huber}]{Prigozhin2016}
Prigozhin, G., Gendreau, K., Doty, J.~P., {et~al.} 2016,
  \href{http://dx.doi.org/10.1117/12.2231718}{in Space Telescopes and
  Instrumentation 2016: Ultraviolet to Gamma Ray, ed. J.-W.~A. den Herder,
  T.~Takahashi, \& M.~Bautz, Vol. 9905}, International Society for Optics and
  Photonics (SPIE), 436

\bibitem[{{Reig} \& {Nespoli}(2013)}]{Reig+13}
{Reig}, P., \& {Nespoli}, E. 2013,
  \href{http://dx.doi.org/10.1051/0004-6361/201219806}{\JournalTitle{\aap},
  551, A1}

\bibitem[{{Remillard} {et~al.}(2022){Remillard}, {Loewenstein}, {Steiner},
  {Prigozhin}, {LaMarr}, {Enoto}, {Gendreau}, {Arzoumanian}, {Markwardt},
  {Basak}, {Stevens}, {Ray}, {Altamirano}, \& {Buisson}}]{Remillard21}
{Remillard}, R.~A., {Loewenstein}, M., {Steiner}, J.~F., {et~al.} 2022,
  \href{http://dx.doi.org/10.3847/1538-3881/ac4ae6}{\JournalTitle{\aj}, 163,
  130}

\bibitem[{{Scott} {et~al.}(1997){Scott}, {Finger}, {Wilson}, {Koh}, {Prince},
  {Vaughan}, \& {Chakrabarty}}]{Scott97}
{Scott}, D.~M., {Finger}, M.~H., {Wilson}, R.~B., {et~al.} 1997,
  \href{http://dx.doi.org/10.1086/304740}{\JournalTitle{\apj}, 488, 831}

\bibitem[{{Sharma} {et~al.}(2022){Sharma}, {Jain}, \& {Dutta}}]{Sharma22}
{Sharma}, P., {Jain}, C., \& {Dutta}, A. 2022,
  \href{http://dx.doi.org/10.1093/mnrasl/slac041}{\JournalTitle{\mnras}, 513,
  L94}

\bibitem[{{Shaw} {et~al.}(2009){Shaw}, {Hill}, {Kuulkers}, {Brandt},
  {Chenevez}, \& {Kretschmar}}]{Shaw09}
{Shaw}, S.~E., {Hill}, A.~B., {Kuulkers}, E., {et~al.} 2009,
  \href{http://dx.doi.org/10.1111/j.1365-2966.2008.14212.x}{\JournalTitle{\mnras},
  393, 419}

\bibitem[{{Snios} {et~al.}(2020){Snios}, {Siemiginowska}, {Sobolewska},
  {Cheung}, {Kashyap}, {Migliori}, {Schwartz}, {Stawarz}, \&
  {Worrall}}]{Snios2020}
{Snios}, B., {Siemiginowska}, A., {Sobolewska}, M., {et~al.} 2020,
  \href{http://dx.doi.org/10.3847/1538-4357/aba2ca}{\JournalTitle{\apj}, 899,
  127}

\bibitem[{{Staubert} {et~al.}(2019){Staubert}, {Tr{\"u}mper}, {Kendziorra},
  {Klochkov}, {Postnov}, {Kretschmar}, {Pottschmidt}, {Haberl}, {Rothschild},
  {Santangelo}, {Wilms}, {Kreykenbohm}, \& {F{\"u}rst}}]{Staubert19}
{Staubert}, R., {Tr{\"u}mper}, J., {Kendziorra}, E., {et~al.} 2019,
  \href{http://dx.doi.org/10.1051/0004-6361/201834479}{\JournalTitle{\aap},
  622, A61}

\bibitem[{{Tanaka}(1986)}]{Tanaka86}
{Tanaka}, Y. 1986, \href{http://dx.doi.org/10.1007/3-540-16764-1\_12}{in IAU
  Colloq. 89: Radiation Hydrodynamics in Stars and Compact Objects, ed.
  D.~{Mihalas} \& K.-H.~A. {Winkler}, Vol. 255}, 198

\bibitem[{{Thalhammer} {et~al.}(2021){Thalhammer}, {Bissinger}, {Ballhausen},
  {Pottschmidt}, {Wolff}, {Stierhof}, {Sokolova-Lapa}, {F{\"u}rst},
  {Malacaria}, {Gottlieb}, {Marcu-Cheatham}, {Becker}, \&
  {Wilms}}]{Thalhammer21}
{Thalhammer}, P., {Bissinger}, M., {Ballhausen}, R., {et~al.} 2021,
  \href{http://dx.doi.org/10.1051/0004-6361/202140582}{\JournalTitle{\aap},
  656, A105}

\bibitem[{{Titarchuk}(1994)}]{Titarchuk94}
{Titarchuk}, L. 1994,
  \href{http://dx.doi.org/10.1086/174760}{\JournalTitle{\apj}, 434, 570}

\bibitem[{{Tsygankov} {et~al.}(2010){Tsygankov}, {Lutovinov}, \&
  {Serber}}]{Tsygankov+10}
{Tsygankov}, S.~S., {Lutovinov}, A.~A., \& {Serber}, A.~V. 2010,
  \href{http://dx.doi.org/10.1111/j.1365-2966.2009.15791.x}{\JournalTitle{\mnras},
  401, 1628}

\bibitem[{{Wang} {et~al.}(2022){Wang}, {Kong}, {Zhang}, {Doroshenko},
  {Santangelo}, {Ji}, {Yorgancioglu}, {Chen}, {Zhang}, {Qu}, {Ge}, {Li},
  {Chang}, {Tao}, {Peng}, \& {Shui}}]{Wang22}
{Wang}, P.~J., {Kong}, L.~D., {Zhang}, S., {et~al.} 2022,
  \href{http://dx.doi.org/10.3847/1538-4357/ac8230}{\JournalTitle{\apj}, 935,
  125}

\bibitem[{{Wilms} {et~al.}(2000){Wilms}, {Allen}, \& {McCray}}]{Wilms00}
{Wilms}, J., {Allen}, A., \& {McCray}, R. 2000,
  \href{http://dx.doi.org/10.1086/317016}{\JournalTitle{\apj}, 542, 914}

\bibitem[{{Wilson} {et~al.}(1995){Wilson}, {Zhang}, {Finger}, {Wilson},
  {Scott}, {Koh}, {Chakrabarty}, {Vaughan}, \& {Prince}}]{Wilson+95}
{Wilson}, C.~A., {Zhang}, S.~N., {Finger}, M.~H., {et~al.} 1995,
  \JournalTitle{\iaucirc}, 6238

\bibitem[{{Wilson-Hodge} {et~al.}(2018){Wilson-Hodge}, {Malacaria}, {Jenke},
  {Jaisawal}, {Kerr}, {Wolff}, {Arzoumanian}, {Chakrabarty}, {Doty},
  {Gendreau}, {Guillot}, {Ho}, {LaMarr}, {Markwardt}, {{\"O}zel}, {Prigozhin},
  {Ray}, {Ramos-Lerate}, {Remillard}, {Strohmayer}, {Vezie}, {Wood}, \& {NICER
  Science Team}}]{Wilson-Hodge18}
{Wilson-Hodge}, C.~A., {Malacaria}, C., {Jenke}, P.~A., {et~al.} 2018,
  \href{http://dx.doi.org/10.3847/1538-4357/aace60}{\JournalTitle{\apj}, 863,
  9}

\bibitem[{{Wolff} {et~al.}(2016){Wolff}, {Becker}, {Gottlieb}, {F{\"u}rst},
  {Hemphill}, {Marcu-Cheatham}, {Pottschmidt}, {Schwarm}, {Wilms}, \&
  {Wood}}]{Wolff16}
{Wolff}, M.~T., {Becker}, P.~A., {Gottlieb}, A.~M., {et~al.} 2016,
  \href{http://dx.doi.org/10.3847/0004-637X/831/2/194}{\JournalTitle{\apj},
  831, 194}

\bibitem[{{Yang} {et~al.}(2018){Yang}, {Zezas}, {Coe}, {Drake}, {Hong},
  {Laycock}, \& {Wik}}]{Yang2018}
{Yang}, J., {Zezas}, A., {Coe}, M.~J., {et~al.} 2018,
  \href{http://dx.doi.org/10.1093/mnrasl/sly085}{\JournalTitle{\mnras}, 479,
  L1}

\end{thebibliography}
\end{document}